 \documentclass{emulateapj}

\usepackage{graphicx}
\usepackage{epstopdf}

\newcommand {\Lya}    {Ly$\alpha$}   

\newcommand {\HI}     {\ion{H}{1}}      
\newcommand {\HeII}   {\ion{He}{2}}   

\newcommand {\OVI}     {\ion{O}{6}}      
\newcommand {\OVII}    {\ion{O}{7}}
\newcommand {\OVIII}   {\ion{O}{8}}

\newcommand {\CII}     {\ion{C}{2}}        
\newcommand {\CIII}    {\ion{C}{3}}        
\newcommand {\CIV}     {\ion{C}{4}}       
\newcommand {\CV}     {\ion{C}{5}}       
\newcommand {\CVI}    {\ion{C}{6}}       

\newcommand {\NV}      {\ion{N}{5}}

\newcommand {\SiVII}   {\ion{Si}{7}}
\newcommand {\SiVI}   {\ion{Si}{6}}
\newcommand {\SiV}   {\ion{Si}{5}}
\newcommand {\SiIV}   {\ion{Si}{4}}
\newcommand {\SiIII}  {\ion{Si}{3}}
\newcommand {\SiII}   {\ion{Si}{2}}

\newcommand {\MgII}    {\ion{Mg}{2}}
\newcommand {\MgI}    {\ion{Mg}{1}}

\newcommand {\CaII}    {\ion{Ca}{2}}

\newcommand {\FeII}   {\ion{Fe}{2}}
\newcommand {\FeI}    {\ion{Fe}{1}}

\newcommand {\kms}    {km~s$^{-1}$}

\newcommand {\FUSE}   {{\it FUSE}} 
\newcommand {\HST}    {{\it HST}}

\newcommand {\etal}   {et~al.} 

\begin{document}

\title{TRACING THE COSMIC METAL EVOLUTION IN THE LOW-REDSHIFT INTERGALACTIC 
MEDIUM\footnote{Based on observations made with the NASA/ESA {\it Hubble Space Telescope}, 
obtained from the data archive at the Space Telescope Science Institute. STScI is operated by the 
Association of Universities for  Research in Astronomy, Inc. under NASA contract NAS5-26555.}  } 

\author{J. Michael Shull$^{a}$, Charles W. Danforth, and Evan M. Tilton}
\affil{ CASA, Department of Astrophysical \& Planetary Sciences, \\
University of Colorado, Boulder, CO 80309, USA \\
\vspace{0.2cm}
$^{a}$also at Institute of Astronomy, University of Cambridge, Cambridge, CB3~OHA, UK
}
 
\email{michael.shull@colorado.edu, danforth@colorado.edu, evan.tilton@colorado.edu }  


\begin{abstract} 

{\small Using the Cosmic Origins Spectrograph aboard the {\it Hubble Space Telescope}, we measured the
abundances of six ions (\CIII, \CIV, \SiIII, \SiIV, \NV, \OVI) in the low-redshift ($z \leq 0.4$) intergalactic 
medium (IGM) and explored C and Si ionization corrections from adjacent ion stages.  Both \CIV\ and 
\SiIV\ have increased in abundance by a factor of $\sim10$ from $z \approx 5.5$ to the present.  We derive ion 
mass densities, $\rho_{\rm ion} \equiv \Omega_{\rm ion} \rho_{\rm cr}$, with $\Omega_{\rm ion}$ expressed 
relative to closure density.  Our models of the mass-abundance ratios (\SiIII/\SiIV) $= 0.67^{+0.35}_{-0.19}$,  
(\CIII/\CIV) $= 0.70^{+0.43}_{-0.20}$, and 
$(\Omega_{\rm CIII} + \Omega_{\rm CIV}) / (\Omega_{\rm SiIII} + \Omega_{\rm SiIV}) = 4.9^{+2.2}_{-1.1}$
are consistent with a hydrogen photoionization rate $\Gamma_{\rm H} = (8\pm2)\times10^{-14}$ s$^{-1}$ at
$z < 0.4$ and specific intensity 
$I_0 = (3 \pm 1) \times 10^{-23} \; {\rm erg~cm}^{-2}~{\rm s}^{-1}~{\rm Hz}^{-1}~{\rm sr}^{-1}$ at the Lyman limit.
We find mean photoionization parameter $\log U = -1.5\pm0.4$, baryon overdensity $\Delta_b \approx 200\pm50$, 
and Si/C enhanced to three times its solar ratio (enhancement of alpha-process elements).  We compare these 
metal abundances to the expected IGM enrichment and abundances in higher photoionized states of carbon (\CV) 
and silicon (\SiV, \SiVI, \SiVII).  Our ionization modeling infers IGM metal densities of 
$(5.4 \pm 0.5) \times 10^5~M_{\odot}~{\rm Mpc}^{-3}$ in the photoionized \Lya\ forest traced by the C and Si ions 
and $(9.1 \pm 0.6) \times 10^5~M_{\odot}~{\rm Mpc}^{-3}$ in hotter gas traced by \OVI.   Combining both phases,
the heavy elements in the IGM have mass density $\rho_Z = (1.5 \pm 0.8) \times 10^6~M_{\odot}~{\rm Mpc}^{-3}$ 
or $\Omega_Z  \approx 10^{-5}$.  This represents $10\pm5$\% of the metals produced by 
$(6 \pm 2) \times 10^8~M_{\odot}~{\rm Mpc}^{-3}$ of integrated star formation with yield $y_m = 0.025\pm0.010$.  
The missing metals at low redshift may reside within galaxies and in undetected ionized gas in galaxy halos 
and circumgalactic medium. }

\end{abstract} 


\section{INTRODUCTION}

It is still uncertain whether a ``missing metals" problem exists for galaxies and the intergalactic medium
(IGM).  The problem was originally proposed  (Pettini 1999) as  a mismatch between the expected 
production rate of heavy elements (metals) and the observed metallicities in galaxies and quasar 
absorption-line systems.   The metal production rate, $\dot{\rho}_Z = y_m \dot{\rho}_*$, can be related to 
$\dot{\rho}_*$, the star formation rate (SFR) density, times the metal yield $y_m$, estimated to lie in the 
range 0.016-0.048 depending on the stellar initial mass function (IMF) and nucleosynthetic yields of massive 
stars and supernovae (Madau \& Dickinson 2014).  The majority of the heavy elements produced by star formation 
in the last 10-13 Gyr were undetected in damped  \Lya\ absorbers (DLAs;  Wolfe \etal\ 2003) or Lyman-break 
galaxies (LBGs, Steidel \etal\ 1999).  Several solutions were proposed for this missing-metals problem, including
the possibility that  heavy elements are stored in gaseous halos or expelled to the IGM by galaxy winds (Ferrara 
\etal\ 2005; Bouch\'e \etal\ 2007) or into the circumgalactic medium (CGM) as suggested by Tumlinson \etal\ (2011).  

Early tests of the IGM hypothesis measured the metallicity at redshifts $z \approx 2-4$ (Songaila 2001, 2005; 
Pettini \etal\ 2003; Boksenberg \etal\ 2003;  Schaye \etal\ 2003; Aguirre \etal\ 2004) using  quasar absorption lines 
of \CIV\  ($\lambda\lambda 1548,1551$) and \SiIV\ ($\lambda\lambda 1393,1403$).  These rest-frame ultraviolet 
(UV) lines are observable with ground-based telescopes for absorption systems that shift into the optical band at 
$z > 1.5$.  Shorter-wavelength transitions of \OVI\ ($\lambda\lambda1032,1038)$, \SiIII\ ($\lambda1206$), and \CIII\  
($\lambda977$) are shifted into the optical at higher redshifts, but confusion with the strong \Lya\ forest complicates 
their measurement.  Metal absorption-line studies have now been extended to larger redshifts $z \approx 5-6$ 
(Simcoe 2006, 2011; Becker \etal\ 2009; Ryan-Weber \etal\ 2006, 2009) with three recent studies (Simcoe \etal\ 2011; 
D'Odorico \etal\ 2013; Cooksey \etal\ 2013) superseding prior work at high redshift.  At  $z < 1.5$, most diagnostic 
absorption lines fall in the ultraviolet and require observations from space with the {\it Hubble Space Telescope} 
(\HST) and {\it Far Ultraviolet Spectroscopic Explorer} (Penton \etal\ 2000, 2004; Danforth \& Shull 2005, 2008;  
Cooksey \etal\  2010).  

In this paper, we analyze the metal-line results obtained from our recent survey of the low-redshift IGM 
with the Cosmic Origins Spectrograph (COS) on \HST\ (Danforth \etal\ 2014).   Observing at low redshifts
avoids the complications of the \Lya\ forest, probes the absorbers directly in the rest-frame UV, and provides access
to a wider range of ionization states.  The current COS survey provides the largest sample of low-$z$ IGM absorbers 
to date, with a total redshift pathlength $\Delta z  \approx 20$ in \HI.  
In Section 2 we describe the portion of the survey relevant to the column density distributions and mass densities of 
six metal ions, for which we have accumulated a substantial number of absorbers.   Our results are described in
Section 3, where we derive low-$z$ mass densities, $\Omega_{\rm ion}$, for all six ions.  The availability of data 
from adjacent ionization states of the same element  (\CIII, \CIV\ and \SiIII, \SiIV) allows us to estimate photoionization 
conditions and correct for unseen ion states.  Because line-blanketing by the \Lya\ forest is generally not a problem 
($<2$\% at low redshifts) these abundances provide accurate values of nucleosynthetically interesting abundance 
ratios (O/C, O/N, Si/C).  In Section~4, we summarize the low-$z$ metallicity and its relevance to SFR history, and we
assess whether a missing-metals issue still exists.

\section{OBSERVATIONAL DATA }  

We have observed six ions (\CIII, \CIV, \NV, \OVI, \SiIII\, \SiIV) in the low-redshift IGM, using the \HST\ Cosmic Origins 
Spectrograph (Green \etal\ 2012) in the medium-resolution mode ($R \approx 18,000$, $\Delta v \approx 17$~\kms) 
with the G130M and G160M gratings that cover 1135--1796~\AA.  Complete results of our IGM survey in the far-UV 
are given in Danforth \etal\ (2014), a study of 75 AGN sight lines that detected over 2500 \HI\ absorption systems
accompanied by more than 350 metal-line systems.   Table 1 provides details on the ions, the number of  metal-ion 
absorbers, their atomic data, and the accessible redshift ranges covered by the  G130M and G160M gratings. \\

To obtain gas-phase IGM metal abundances,  we follow the standard technique of measuring absorption lines of the
UV resonance transitions of key ion species and making ionization corrections.  Most metallicity surveys are made
with one or both lines of the doublet transitions of \CIV\  ({\it 2s-2p} 1548.204, 1550.781~\AA) and \SiIV\ ({\it 3s-3p} 
1393.760, 1402.773~\AA) owing to their line strength and accessibility longward of the \Lya\ forest 
($\lambda_{\rm rest} > 1215.67$~\AA). Although line-blanketing by the  \Lya\ forest becomes strong at high redshift, 
our COS survey has access to the doublet transition of \OVI\ ({\it 2s-2p} 1031.926, 1037.617~\AA) and single-line absorption 
from strong transitions of \CIII\ ({\it 2s$^2$-2s\,2p} 977.020~\AA) and \SiIII\ ({\it 3s$^2$-3s\,3p} 1206.500~\AA).  This 
list is supplemented by the \NV\ doublet ({\it 2s-2p} 1238.82, 1242.804~\AA).    
The \CIII\ and \CIV\ absorbers are not present in the same gas, since \CIII\ $\lambda 977$ shifts into the G130M band
at $ z > 0.162$ and \CIV\ $\lambda 1548$ shifts out of the G160M band at  $z > 0.160$.  Thus, the ion-ratio comparisons
with COS data are statistical.   We have also examined the \CIII / \CIV\ ratio in our STIS + FUSE survey (Tilton \etal\ 2012) 
where both lines are present in the same absorbers;  we find good agreement with the COS results.

\section{RESULTS AND INTERPRETATION}  

\subsection{Metal Ion Abundances in the IGM} 

We express the mass density of a metal ion, $\rho_{\rm ion} = \Omega_{\rm ion} \rho_{\rm cr}$, relative to $\rho_{\rm cr}$, 
the cosmological closure density.  For Hubble constant $H_0 = (70~{\rm km~s}^{-1}~{\rm Mpc}^{-1}) h_{70}$, we can write
$\rho_{\rm cr} = (3 H_0^2/8 \pi G) = 9.205 \times 10^{-30}~h_{70}^2~{\rm g~cm}^{-3}$, or in more convenient galactic units, 
$1.36 \times 10^{11}~h_{70}^2 \, M_{\odot}~{\rm Mpc}^{-3}$.  The mean comoving baryon overdensity (at $z = 0$) is written
$\bar{\rho}_b = [\Omega_b \rho_{\rm cr} (1-Y_p)/m_H] \approx 1.90 \times 10^{-7}$~cm$^{-3}$, appropriate for 
$\Omega_b = 0.0461$ (Hinshaw \etal\ 2013) and primordial helium abundance $Y_p = 0.2485$ (Aver \etal\ 2013). 
The ion mass density is found by integrating over its observed column density distribution function (CDDF) of absorbers, 
$f(N,z) \equiv (\partial ^2 {\cal N} / \partial N \, \partial z)$, as described in Tilton \etal\ (2012),
\begin{eqnarray}
    \Omega_{\rm ion} &=& \frac { H_0 \, m_{\rm ion}} {c \, \rho_{\rm cr}} \int_{\rm N_{\rm min}}^{\rm N_{\rm max}} 
           f(N,z) \,  N \,  dN     \nonumber \\
        &=&  (1.365 \times 10^{23}~{\rm cm}^2) \, h_{70}^{-1} \left( \frac {m_{\rm ion}} {{\rm amu}} \right)
           \int_{\rm N_{\rm min}}^{\rm N_{\rm max}}  f(N,z) \,  N \,  dN \; .
\end{eqnarray}  
The integral is performed as a sum over logarithmic bins of size $\Delta \log N \approx 0.2$.  The range of integration is set 
at the lower end ($N_{\rm min}$) by our 30~m\AA\ equivalent width threshold and at the upper end ($N_{\rm max}$) by 
the column density where absorber statistics become uncertain.   Depending on the slope of the distribution, 
$f(N,z) \propto N^{-\beta}$, we can make corrections for absorption systems outside this range.  Appendix~A describes our 
formalism for fitting these distributions and integrating power-law fits to $f(N,z)$.   This technique provides a check 
on the numerical integration and allows us to assess the fiducial metal content, extrapolated to column densities above and 
below the observed range.  \\

Table 2 gives the $\Omega_{\rm ion}$ values for all six ions, measured in low-$z$ surveys with HST/STIS (Tilton \etal\ 
2012) and HST/COS (Danforth \etal\ 2014).  Although our metal absorber list is large, there is considerable overlap
with prior studies (Danforth \& Shull 2008; Cooksey \etal\ 2011; Tilton \etal\ 2012).  We express values of $\Omega_{\rm ion}$ 
in units of $10^{-8} h_{70}^{-1}$.  All values from our previous STIS survey were recalculated from the Tilton \etal\ (2012) 
absorber lists to match the column density ranges in Table 2.  We then compare the low-$z$ measurements to 
values from previous surveys, including Songaila (2001, 2005), Becker \etal\  (2009), Simcoe (2006), Simcoe \etal\ (2011), 
and Ryan-Weber \etal\ (2006).  All values of $\Omega_{\rm ion}$ were revised to agree with our adopted ($\Lambda$CDM) 
cosmological parameters, $h = 0.7$, $\Omega_m = 0.275$, $\Omega_{\Lambda} = 0.725$.   \\

The distributions in column density were shown in Danforth \etal\ (2014) plotted versus column density, $N_{\rm ion}$, for 
\CIV, \SiIV, \CIII, \SiIII, and \OVI.   These distributions illustrate the systematic uncertainties inherent in computing values of 
$\Omega_{\rm ion}$ from Equation (1) over the measured range of column densities
($\log N_{\rm min} < \log N < \log N_{\rm max}$). For several ions, the discrete sums are sensitive to the small number of 
absorbers at the high end, as well as the reduced pathlength $\Delta z$ for absorbers at the low end.   As was done by previous 
authors (Cooksey \etal\ 2010, 2013; D'Odorico \etal\ 2013) we fit power laws to the CDDF and extrapolate to a standard 
range of column densities as discussed in Appendix A and Table~3.   These distributions are often more useful to observers 
than the single parameter $\Omega_{\rm ion}$, and they convey information about the sparse statistics for metal 
absorbers at both low and high column densities.    \\

Figure 1 shows the evolution of $\Omega_{\rm CIV}$ with redshift.  The \CIV\ data include high-redshift measurements 
($z \approx 4.5-6$) by Pettini \etal\  (2003), Becker \etal\ (2009), Ryan-Weber \etal\ (2009), Simcoe \etal\ 2011), and 
D'Odorico \etal\ (2013).  We corrected the Pettini results for our values of $H_0$ and $\Omega_m$ using the scaling 
$\Omega_{\rm ion} \propto \Omega_m^{1/2} h^{-1}$.  Values at intermediate redshifts ($z \approx 1.6-4.5$) are taken 
from surveys by Boksenberg \etal\ (2003), D'Odorico \etal\ (2010), and Boksenberg \& Sargent (2014).  The latter paper
includes updated values for $\Omega_{\rm CIV}$ provided by A. Boksenberg (2014, private communication). 
The data at $z < 1$ come from the HST survey by Cooksey \etal\ (2010) reported in two redshift bins:
$\Omega_{\rm CIV} = 6.24^{+2.88}_{-2.14}$ ($10^{-8} h_{70}^{-1}$) from 17 \CIV\ absorbers at $\langle z \rangle = 0.383$ 
and $\Omega_{\rm CIV} = 6.35^{+2.52}_{-1.99}$($10^{-8} h_{70}^{-1}$) from 19 \CIV\ absorbers at $\langle z \rangle = 0.786$.
For the entire redshift sample, they found $\Omega_{\rm CIV} = 6.20^{+1.82}_{-1.52}$ ($10^{-8} h_{70}^{-1}$) from 36 \CIV\ 
absorbers at $\langle z \rangle = 0.654$.  The lowest redshift points at $\langle z \rangle < 0.1$ come from the Colorado 
group's surveys:  
$\Omega_{\rm CIV} = 7.7 \pm 1.5$ ($10^{-8} h_{70}^{-1}$) from 24 \CIV\ absorbers (Danforth \& Shull 2008), 
$\Omega_{\rm CIV} = 8.1^{+4.6}_{-1.7}$ ($10^{-8} h_{70}^{-1}$) from 29 \CIV\ absorbers (Tilton \etal\ 2012),
and $\Omega_{\rm CIV} = 10.1^{+5.6}_{-2.4}$ ($10^{-8} h_{70}^{-1}$) from 49 \CIV\ absorbers (Danforth \etal\ 2014). \\
 
Figure 2 shows the evolution of  $\Omega_{\rm SiIV}$  with redshift, including high-redshift measurements ($z \approx 2-5$) 
by Songaila (2001, 2005) converted to our parameters ($H_0$ and $\Omega_m$).  We also show data at $z \approx 2.4$ 
(Scannapieco \etal\ 2006), at $z \approx 3.3$ (Aguirre \etal\ 2004), and at $z = 1.9-4.5$ (Boksenberg \& Sargent 2014).  
The latter paper includes new values for $\Omega_{\rm SiIV}$ provided by A. Boksenberg (2014, private communication.)   
The low-redshift data come from the HST survey by Cooksey \etal\ (2011) at $z < 1$.  The lowest redshift points at 
$\langle z \rangle < 0.1$ are based on the Colorado group's recent surveys:  
$\Omega_{\rm SiIV} = 4.5^{+3.0}_{-1.2}$ ($10^{-8} h_{70}^{-1}$) from 30 \SiIV\ absorbers (Tilton \etal\ 2012)
and $\Omega_{\rm SiIV} = 2.1^{+1.0}_{-0.5}$ ($10^{-8} h_{70}^{-1}$) from 31  \SiIV\ absorbers (Danforth \etal\ 2014). 
Evidently, the \SiIV\  measurements still have some uncertainty arising from the column density distribution function.

\subsection{Metagalactic Ionizing Background}   

In Section 3.3 we will analyze the ionization states of C, Si, and O probed by the observed ions.  This analysis 
requires photoionization modeling using estimates of the metagalactic flux of ionizing photons.   Through analytic 
formulae, we can relate the hydrogen photoionization rate $\Gamma_{\rm H}$ for optically thin absorbers to the specific 
intensity $I_{\nu}$ and its value $I_0$ at the Lyman limit (912~\AA), as well as to $\Phi_0$, the one-sided (unidirectional) 
flux of ionizing photons.   Because there has been some debate in the modeling literature over the strength 
of the metagalactic ionizing radiation field,  we begin with clear definitions of flux and intensity.   Part of the confusion 
arises from geometric differences between an isotropic radiation field, appropriate for intergalactic space far from 
individual sources, and a unidirectional radiation field arising from a single nearby source.   In models for the 
local-source case, all photons enter the absorber at normal incidence from one side.   In the isotropic case, half the 
photons are directed into the forward hemisphere and half backward.  One must also account for the angular dependence 
of the radiation, with $\langle \cos \theta \rangle = 1/2$ averaged over the forward hemisphere.  Thus, for an isotropic 
radiation field, the unidirectional flux $\Phi_0 = n_{\gamma} (c/4)$, and for the local source $\Phi_0 = n_{\gamma} c$,
where $n_{\gamma}$ is the number density of ionizing photons.  Geometric factors are less critical for optically thin 
absorber models, but one must be careful when using flux or intensity parameters to derive $n_{\gamma}$, the 
photoionization parameter $U = n_{\gamma} / n_{\rm H}$, and the ionization fractions of H, He, and heavy elements.

The ionizing radiation background remains somewhat uncertain, despite two decades of estimating its strength 
and spectrum (Haardt \& Madau 1996, 2001, 2012; Shull \etal\ 1999; Faucher-Gigu\`ere \etal\ 2009).   Figure 3 shows
several estimates of  the ionizing background in the extreme ultraviolet (EUV) and soft X-ray.  The labels (HM01, HM05, 
HM12,  FG11) refer to published or unpublished tabulations based on earlier publications. The HM05 spectrum refers 
to the 2005 August update to the Haardt \& Madau (2001) radiation field\footnote{Unpublished version of the metagalactic 
Quasars+Galaxies ionizing spectra described in Haardt \& Madau (2001) and provided for inclusion in \textsc{Cloudy}.} 
and the FG11 spectrum refers to the 
2011 December update\footnote{Retrieved from \url{http://galaxies.northwestern.edu/uvb/}}
to the spectrum in Faucher-Gigu\`ere \etal\ (2009). The PL spectrum refers to our constructed  broken power-law 
distribution consistent with recent \HST/COS composite spectrum (Shull \etal\ 2012) in the EUV (500-912~\AA) and 
connected to soft X-ray observations in the Lockman Hole (Hasinger 1994).  The hard X-ray background 
(Churazov \etal\ 2007) from 10--100 keV has been included in some backgrounds (HM01, HM05, HM12);   these
energetic photons have little effect on the ionization state of H, He, and heavy elements.   

The hydrogen-ionizing background  $\Phi_0$ and photoionization rate $\Gamma_{\rm H}$  of HM12 have been questioned 
(Kollmeier \etal\ 2014) as inconsistent with the column density distribution of low-redshift \Lya\ (\HI) absorbers (Danforth
\etal\ 2014).  The numerical simulations in their analysis were based on the HM12 background, which was deemed too low 
by a factor of 5.  However, they noted better agreement when the HM01 background was used.  A similar comparison 
(Egan \etal\ 2014) also obtained good agreement with the \Lya\ distribution, using our group's grid-code (\texttt{Enzo}) 
simulations (Smith \etal\ 2011) based on the HM01 ionizing background.  The primary difference between HM01 and HM12 
radiation fields comes from different assumptions about the escape fraction ($f_{\rm esc}$) of Lyman continuum (LyC) radiation 
from galaxies (Shull \etal\ 1999).  In the  HM12 formulation, the ionizing background is dominated by AGN, the escape fraction 
is parameterized as $f_{\rm esc}(z) \approx (1.8 \times 10^{-4}) (1+z)^{3.4}$,  and galaxies make essentially no contribution to 
the photoionization rate at $z < 1$.  Such low values of $f_{\rm esc}$ are probably unrealistic, since the usual observational 
constraints (direct LyC detection, diffuse H$\alpha$ emission, and metal ion ratios) depend on small-number statistics and 
geometric uncertainties.   If the LyC photons escape through vertical cones above star-forming regions in galactic disks 
(Dove \& Shull 1994), the observational constraints require large samples to deal with inclination bias.  High-redshift surveys 
(Shapley \etal\ 2006) find either large transmitted LyC fluxes or none at all.    As a consequence, if $f_{\rm esc}$ were only 5\%, 
one would need to survey well over 20 galaxies to obtain a statistically accurate escape fraction. 

Systematic uncertainties also affect  indirect limits on the metagalactic background from diffuse H$\alpha$ emission (Stocke 
\etal\ 1991; Donahue \etal\ 1995; Vogel \etal\ 1995; Weymann \etal\ 2001).  Translating H$\alpha$ surface brightnesses into 
limits on $\Phi_0$ or $\Gamma_{\rm H}$ requires assumptions about emitting geometries and radiative transfer of LyC.   
Adams \etal\ (2011) used integral-field spectroscopy to set deep limits on H$\alpha$ emission from the outskirts of two
low surface brightness, edge-on Sd galaxies, UGC~7321 and UGC~1281. 
They constrained the  \HI\ photoionization rate to be $\Gamma_{\rm H} < 2.3 \times 10^{-14}$~s$^{-1}$ (UGC~7321) and 
$\Gamma_{\rm H} < 14 \times 10^{-14}$~s$^{-1}$ (UGC~1281).  The UGC~7321 limit comes from a ($5\sigma$) absence
of H$\alpha$, and it lies well below previous observational limits and theoretical predictions.  However, we are reluctant 
to adopt such a low value of $\Gamma_{\rm H}$ or $\Phi_0$ from a single object\footnote{Their stringent limit comes from 
UGC~7321, whereas the limit toward UGC~1281 is a factor of 6 higher.  In a subsequent AAS abstract (Uson \etal\ 2012) the 
same authors claim to have detected H$\alpha$ emission toward UGC~7321, with an inferred 
$\Gamma_H = 3.4 \times 10^{-14}$~s$^{-1}$ that is larger than their 2011 upper limit.}.  Their method of inferring the 
metagalactic flux $\Phi_0$ and ionization rate $\Gamma_{\rm H}$ relies on a large and uncertain geometric correction for 
the gas cloud aspect ratio, $A_{\rm tot} / A_{\rm proj}$, of total exposed area to projected area.  As noted by Stocke \etal\ (2001), 
this ratio is 4 for a sphere and $\pi$ for a cylinder viewed transversely.   The value
$\langle A_{\rm tot} / A_{\rm proj} \rangle = 24.8^{+3.4}_{-1.5}$ adopted by Adams \etal\ (2011) is based on the projection of 
a  homogeneous thin disk viewed nearly edge-on, whereas the H$\alpha$ emission likely arises in smaller clumps of gas, 
whose geometries are probably closer to spheres or cylinders.   With the inferred ionizing flux scaling as
$\Phi_0 \propto (A_{\rm tot} / A_{\rm proj})^{-1}$, their lower limit arises from a 6-8 times larger aspect ratio compared to 
values for spherical or cylindrical clouds.   \\

In view of the extensive experimental and theoretical literature on metagalactic fluxes, hydrogen ionization rates, and IGM 
ionization conditions, the limits on low-redshift ionizing radiation remain uncertain.   In Section 3.3, we use ratios of adjacent
ion stages (\CIII/\CIV\ and \SiIII/\SiIV) from  IGM absorption-line spectroscopy to constrain 
$\Phi_0 \approx 10^4$~cm$^{-2}$~s$^{-1}$.  A similar technique, based on the ions  \FeI/\FeII\ and \MgI/\MgII\  in the 
low-redshift absorption system  around NGC~3067 probed by the background QSO 3C~232, gave a lower limit 
$\Phi_0 > 2600$~cm$^{-2}$~s$^{-1}$.  Our models in Section 3.3 use $\Phi_0 \approx 10^4$ cm$^{-2}$~s$^{-1}$, 
$\Gamma_{\rm H} \approx (8\pm2) \times 10^{-14}$ s$^{-1}$, and 
$I_0 \approx (3\pm1) \times 10^{-23}~{\rm erg~cm}^{-2}~{\rm s}^{-1}~{\rm Hz}^{-1}~{\rm sr}^{-1}$.  These parameters are in 
reasonable agreement with past calculations, observational estimates (other than the single galaxy probed by Adams
\etal\ 2011), and with our low-redshift absorber data on \CIII, \CIV, \SiIII,  and \SiIV.   We now explore this issue quantitatively.  \\

For an isotropic radiation field of specific intensity $I_{\nu}$, the normally incident photon flux per frequency is $(\pi I_{\nu}/h \nu)$ 
into an angle-averaged, forward-directed effective solid angle of $\pi$ steradians.  The isotropic photon flux striking an atom or 
ion is $4 \pi (I_{\nu} / h \nu)$.  The hydrogen photoionization rate follows by integrating this photon flux times the photoionization 
cross section over frequency from threshold ($\nu_0$) to $\infty$.  
\begin{eqnarray} 
  \Gamma_{\rm H} &=&  \int _{\nu_0}^{\infty} \frac {4 \pi \, I_{\nu}}{h \nu} \sigma_{\nu} \, d \nu \approx
      \left( \frac { 4 \pi I_0 \sigma_0} {h}  \right) \int_{1}^{\infty} x^{-\alpha-\beta-1} \, dx  \nonumber \\
       &=&   \frac { 4 \pi I_0 \sigma_0} {h (\alpha+\beta) }   \; .
\end{eqnarray}
Here, we define the dimensionless variable $x = \nu / \nu_0$ and approximate the frequency dependence of
specific intensity and photoionization cross section by power laws, $ I_{\nu} = I_0  (\nu / \nu_0)^{-\alpha}$ and
$\sigma_{\nu}  =  \sigma_0 (\nu / \nu_0)^{-\beta}$, where $\sigma_0 = 6.30 \times 10^{-18}$~cm$^2$ and 
$\beta \approx 3$.  The integrated unidirectional flux of ionizing photons is
\begin{equation}
   \Phi_0 = \int_{\nu_0}^{\infty} \frac { \pi I_{\nu} } {h \nu} \; d \nu  =
    \left( \frac { \pi I_0 } {h}  \right) \int_{1}^{\infty} x^{-\alpha - 1} \, dx = 
    \frac {\pi I_0} {h \alpha} \;  \; ,
\end{equation} 
which can be related to the density of hydrogen-ionizing photons by $\Phi_0 = n_{\gamma} (c/4)$.  We note
that this definition differs by a factor of 1/4 from that in \textsc{Cloudy}, which adopts the convention that
$\Phi_0 = n_{\gamma} c$, appropriate for all photons arriving at normal incidence from one direction.  We then 
use $I_0 = (h \alpha / \pi) \Phi_0$ to relate $\Gamma_{\rm H}$ to the flux and spectrum parameters:
\begin{eqnarray}
   \Gamma_{\rm H} &=& 4 \sigma_0 \Phi_0 \left(  \frac {\alpha} { \alpha + \beta } \right) = 
           (8.06 \times 10^{-14}~{\rm s}^{-1}) \Phi_4   \\
   \Gamma_{\rm H} &=&   \left( \frac { 4 \pi I_0 \sigma_0 } { \alpha + \beta } \right) = 
           (2.71\times10^{-14}~{\rm s}^{-1}) I_{-23}  \; \; .
\end{eqnarray}
We will see that $\Phi_4 \approx 1$ and $I_{-23} \approx 3$ give reasonable fits to the ionization ratios, 
where we scale the incident flux of ionizing photons and specific intensity to characteristic values, 
$\Phi_0 = (10^4~{\rm cm}^{-2} \,  {\rm s}^{-1}) \Phi_4$
and $I_0 = (10^{-23} \; {\rm erg~cm}^{-2}~{\rm s}^{-1}~{\rm Hz}^{-1}~{\rm sr}^{-1}) I_{-23}$ at the hydrogen Lyman limit 
($h \nu_0 = 13.60$~eV). We adopt $\beta \approx 3$ for hydrogen and an AGN composite spectrum (Shull \etal\ 2012) 
with mean spectral index $\alpha \approx 1.41$ between $1.0-1.5$~ryd.   For these parameters, we can write the 
photoionization parameter $U = n_{\gamma} / n_H$ as,
\begin{equation}
   U \equiv \left( \frac {4 \Phi_0} {n_H c} \right) = (7.02 \times 10^{-2}) \Phi_4 \Delta_{100}^{-1}  \; \; ,
\end{equation}  
where the absorber hydrogen density is $n_H = (1.90 \times 10^{-7}~{\rm cm}^{-3}) \Delta_b$ at $z = 0$ with
baryon overdensity scaled to $\Delta_b \equiv 100 \Delta_{100}$.  

Recent calculations of the metagalactic ionizing background (Haardt \& Madau 2012) yield a hydrogen 
photoionization rate, which can be fitted to $\Gamma_H = (2.28 \times 10^{-14}~{\rm s}^{-1}) (1+z)^{4.4}$
over the range $0 \leq z \leq 0.7$.  This HM12 ionization rate at $z =0$ would correspond to 
$\Phi_0 = [\Gamma_{\rm H}  (\alpha + \beta)/ 4 \sigma_0 \, \alpha] \approx 2630~{\rm cm}^{-2}~{\rm s}^{-1}$ and
$I_{-23} \approx 0.823$ for the radio-quiet AGN spectral index ($\alpha = 1.57$) assumed by HM12.  These
fluxes are lower by a factor of 3 compared to the $z = 0$ metagalactic radiation fields from AGN and galaxies
calculated by Shull \etal\ (1999), $I_{\rm AGN} = 1.3^{+0.8}_{-0.5} \times 10^{-23}$ and
$I_{\rm Gal} = 1.1^{+1.5}_{-0.7} \times 10^{-23}$, respectively.   Adding these two values with propagated errors
gives a total intensity and hydrogen ionization rate of
$I_{\rm tot} =  2.4^{+1.7}_{-0.9} \times 10^{-23} \; {\rm erg~cm}^{-2}~{\rm s}^{-1}~{\rm Hz}^{-1}~{\rm sr}^{-1}$ and
$\Gamma_{\rm H}  =  6.0^{+4.2}_{-2.1} \times 10^{-14} \; {\rm s}^{-1}$.   These backgrounds are consistent with those 
that we estimate below from the C and Si ionization states.   The difference between these radiation 
fields appears to be the small contribution of galaxies assumed in the HM12 background compared to that of
Shull \etal\ (1999), HM01, and HM05.   For these reasons, as well as the issues raised by Kollmeier \etal\  (2014), 
we normalize our modeled spectral energy distributions (SEDs) to unidirectional, normally incident  fluxes 
$\Phi_0 = 10^4~{\rm cm}^{-2}~{\rm s}^{-1}$, corresponding to specific intensity 
$I_0 \approx 3 \times 10^{-23} \; {\rm erg~cm}^{-2}~{\rm s}^{-1}~{\rm Hz}^{-1}~{\rm sr}^{-1}$ and hydrogen
photoionization rate $\Gamma_{\rm H} \approx 8 \times 10^{-14}$ s$^{-1}$.

\subsection{Metal Densities at Low Redshift}

The IGM metal density traced by these six ion species can be estimated by correcting for unseen ionization states
and scaling to total metal abundances.  For the latter, we initially adopted solar abundances to compute the
fractional contributions of C, Si, or O to the total metal abundances by mass.   However, after comparing the C and Si
abundances, we prefer an abundance pattern in which alpha-process elements (e.g., Si and O) are enhanced
relative to C owing to early nucleosynthesis sources by massive stars.  In high-redshift spectra, observers usually 
measure \CIV\ and  \SiIV, and sometimes \CII\ and \SiII, but they often lack the important intermediate ion states \CIII\ 
and \SiIII, whose absorption lines at 977.0~\AA\  and 1206.5~\AA\ can be confused by the \Lya\ forest.  Here, in our 
low-$z$ survey, we measure \CIII\ and \CIV\ as well as \SiIII\ and \SiIV.   
We use the mean (statistical) values of the ratios of adjacent ion states,  \CIII/\CIV\ and \SiIII/\SiIV,  to constrain the IGM
density and the strength and shape of the ionizing radiation background.  We then use the ionization corrections for 
C and Si to estimate consistent individual metallicities.  By comparing these metallicities to their expected abundance 
ratios, we derive additional constraints on Si/C abundance enhancement, radiation field, and IGM density.  

The mean observed ionization fractions of the ensemble of metal-line absorbers are:
\begin{equation}
       \frac {\Omega_{\rm CIII} } { \Omega_{\rm CIV} }     = 0.70^{+0.43}_{-0.20}  \; \; \; {\rm and} \; \;  \; 
       \frac {\Omega_{\rm SiIII} } { \Omega_{\rm SiIV} }    = 0.67^{+0.35}_{-0.19}    \; \;  .
\end{equation}
We will show that for a range of ionizing background shapes, these ratios are consistent with photoionization parameter 
$\log U \approx -1.5 \pm 0.4$. As noted earlier, $U = n_{\gamma}/n_{\rm H}$ expresses the density ratio of ionizing photons to 
hydrogen, where $n_{\rm H}  = (1.90 \times 10^{-7}$~cm$^{-3}) \Delta_b (1+z)^3$, with baryon overdensity $\Delta_b$.
Because the absorption lines of \CIII\ $\lambda977$ and \CIV\ $\lambda1548$ redshift in and out of the COS G130M and 
G160M bands at $z \approx 0.16$, our COS survey does not observe \CIII\ and \CIV\ in the same absorbers.  Thus, there 
may be some concern that these ions are experiencing different radiation fields, in which the ionization parameter, 
$U(z) \propto (1+z)^{1.4}$ for an ionization rate $\Gamma_{\rm H}  \propto (1+z)^{4.4}$ and hydrogen density 
$n_{\rm H}  \propto (1+z)^3$.  In the COS sample (Danforth \etal\ 2014) the mean redshifts of the carbon absorbers are 
$\langle z_{\rm CIII} \rangle =  0.35$ and $\langle z_{\rm CIV} \rangle =  0.06$, so we might expect a 0.15 dex 
offset in $U$ between those absorbers.  However, we have also examined low-redshift data from our \FUSE\ and 
STIS survey (Tilton \etal\ 2012), in which both ions are both seen in the same absorbers.  In those 28 absorbers,
with $\langle z_{\rm CIII} \rangle =  0.15$ and $\langle z_{\rm CIV} \rangle =  0.056$, the \CIII/\CIV\ ratios are essentially 
the same as in the COS data.   Therefore, we choose to make no correction for possible small offsets in the radiation field.

We explore the ionizing spectra and their effects on the observed ion densities through a series of photoionization 
calculations with version 13.03 of \textsc{Cloudy}, last described by Ferland \etal\ (2013).  All spectra were flux-normalized 
to $\Phi_0 = 10^4 \; {\rm cm}^{-2} \,{\rm s}^{-1}$.  The hydrogen density was related to baryon overdensity $\Delta_b$, and
the metallicity in the simulations was taken as 0.1 solar, consistent with inferences for the mean in the low-$z$ IGM 
(Shull \etal\ 2012).  The input spectra consisted of a variety of SEDs of ionizing photons, reflecting the uncertain intensities 
in the EUV and soft X-ray (Shull \etal\ 2012; Haardt \& Madau 2012;  Faucher-Gigu\`ere \etal\ 2009).  We constrain the SED 
from the observed abundance ratios of adjacent ionization states, \CIII/\CIV\ and \SiIII/\SiIV\ and then estimate the total mass 
densities of C and Si,
\begin{eqnarray}  
    \Omega_{\rm C}  &=&   \left( \Omega_{\rm CIII} +  \Omega_{\rm CIV} \right)  \left( 
         \frac { {\rm C_{\rm tot}} } {{\rm C~III} +  {\rm C~IV} }  \right)    \\
    \Omega_{\rm Si} &=&  \left( \Omega_{\rm SiIII} +  \Omega_{\rm SiIV} \right)  \left(
       \frac {{\rm Si_{\rm tot}} } { {\rm Si~III} +  {\rm Si~IV}  } \right)  \; \; .
\end{eqnarray}
The second parenthetical terms in these relations are the ionization correction factors, abbreviated CF$_{\rm C}$ and 
CF$_{\rm Si}$, which quantify the amount of C and Si in higher ion states.   As discussed more fully in Appendix~B, ionizing
EUV photons can produce helium-like carbon (\CV) by ionizing  \CIV\ at energies $E \geq 64.49$~eV.  The EUV photons also 
make \SiV, \SiVI, \SiVII\ with production threshold energies 33.49 eV (from \SiIII), 45.14 eV (from \SiIV), and 166.77 eV 
(from \SiV) respectively.  These ions can also be produced by inner-shell ionization by soft X-rays  ($E \geq 1.9$~keV) 
followed by Auger electron emission, which boosts the ionization by two or more stages.  Further discussion and analytic 
estimates of the abundances of higher ion states of C and Si are provided in Appendix~B. 

Figure 4 shows the ratios, \CIII/\CIV\ and \SiIII/\SiIV, as functions of $\Delta_b$, and Figure~5 shows
the ionization correction factors CF$_{\rm C}$  and CF$_{\rm Si}$ vs.\ $\Delta_b$ for various SEDs.  Our  \textsc{Cloudy} 
photoionization models of the observed C and Si ion ratios, together with their uncertainties, suggest that 
$\Delta_b \approx 70-230$ for the HM12 and PL spectral distributions.   The agreement is not as good with the FG11 and 
HM05 background, with HM12 and PL models providing better agreement.   
The corresponding correction factors to the two observed stages of C and Si are 
${\rm CF}_{\rm C} = \Omega_{\rm C}  / \left[  \Omega_{\rm CIII} +  \Omega_{\rm CIV} \right] =  2.0^{+1.0}_{-0.5}$ and 
${\rm CF}_{\rm Si} = \Omega_{\rm Si} / \left[  \Omega_{\rm SiIII} +  \Omega_{\rm SiIV} \right] =  6^{+4}_{-3}$.  

An additional constraint comes from the C and Si metallicities inferred from our ionization corrections, which
must be consistent with their relative abundances by mass.  For solar abundances, the Si/C ratio is  
$\rho_{\rm Si} / \rho_{\rm C} = 0.238$ by mass, but our survey suggests that Si must be enhanced by a factor 
of 3 (Figure 6) probably owing to ``alpha-process"  nucleosynthesis.  Using the observed mass densities in Table~2, 
 $(\Omega_{\rm CIII} + \Omega_{\rm CIV}) = 17.2^{+5.9}_{-2.7}$ $(10^{-8} h_{70}^{-1})$ and 
 $(\Omega_{\rm SiIII} + \Omega_{\rm SiIV}) = 3.5^{+1.0}_{-0.5}$ $(10^{-8} h_{70}^{-1})$, we can relate the Si/C 
abundance ratio to the ion correction factors:
\begin{equation}
    \frac { {\rm CF}_{\rm Si} } { {\rm CF}_{\rm C} } = (1.17^{+0.52}_{-0.25} )
         \left[ \frac {\rho_{\rm Si} / \rho_{\rm C} } {0.238} \right]   \;  \;  .
\end{equation} 
From the constraints in Figures 4, 5, and 6, we find that $\Delta_b \approx 200 \pm 50$ for ionizing flux
$\Phi_0 = 10^4$ cm$^{-2}$~s$^{-1}$.  If we were to lower the ionizing background to 
$\Phi_0 = (3000-5000)$ cm$^{-2}$~s$^{-1}$, our ionization models for the C and Si ion ratios would require
lowering the baryon overdensity to $\Delta_b = 50-100$.  As can be seen in Figures 4 and 5, the ionization
correction factors would then become unrealistically large, particularly for silicon which would rise
to ${\rm Si_{\rm tot}} / ({\rm Si~III} +  {\rm Si~IV})  > 100$.   Thus, the observed metal-line systems probably 
reside in higher density regions than \Lya-forest systems at lower column densities.  Therefore, the observed
ionization ratios of \CIII/\CIV\ and \SiIII/\SiIV\ provide additional evidence for a higher metagalactic radiation field, 
$\Phi_0 \approx 10^4$ cm$^{-2}$~s$^{-1}$ and ionization rate $\Gamma_H \approx 8 \times 10^{-14}$ s$^{-1}$.
These values, while still uncertain, are consistent with theoretical calculations (Haardt \& Madau 2001; Shull \etal\ 
1999) and inferences from the \Lya\ column-density distribution (Kollmeier \etal\ 2014; Egan \etal\ 2014).  

Thus, the enhanced Si/C abundances are consistent with the PL, FG11, and HM05 radiation fields, but not with HM12.
However, the constraints are not always in full agreement.   Our choice of $\Delta_b$ is guided primarily 
by the need for sensible (finite and consistent) ionization correction factors for C and Si (Figures~5 and 6).   Our inference 
of enhanced Si/C and softer radiation fields was also based on these consistency arguments.    High radiation fields in the 
EUV and soft X-ray produce far too many high ions (\CV, \SiV, \SiVI, \SiVII) to be consistent with the expected metallicities.   
Because the high ionization states of Si are sensitive to hard-EUV and soft X-ray photons, the Si ionization correction 
curves in Figure 5 are too high for the HM12 radiation field, which has elevated fluxes between 3--300~\AA.  The bottom 
panel of Figure 5 shows that CF$_{\rm Si} > 10$ for $\Delta_b \leq 300$, inconsistent with the values of CF$_{\rm C}$
unless the Si/C abundance is enhanced above solar ratios.  Figure 4 shows that the adjacent-ion ratios for \CIII/\CIV\ and 
\SiIII/\SiIV\  are somewhat inconsistent with the softer radiation backgrounds (FG11 and HM05), preferring the HM12 and 
PL distributions.   Clearly, there is a need for additional work to infer the uncertain metagalactic background between 
2--20 ryd, particularly between the \HeII\ edge (54.4 eV) and the soft X-ray band (0.3--3~keV) responsible for inner-shell 
ionization of many metal ions.  

The total metal abundance, $\Omega_Z$, can be estimated by dividing $\Omega_{\rm C}$,  $\Omega_{\rm Si}$, and
$\Omega_{\rm O}$ by the individual mass fractions of these elements relative to all heavy elements.  From the recent 
solar abundance calculations of Caffau \etal\ (2011), the relevant fractions are 18.2\% (C), 4.34\% (Si), and 44.0\% (O).  
Using our corrections for ionization state and metal fractions of C and Si, we find mean mass densities of 
$\Omega_{\rm C} =  3.4 \times 10^{-7}$ and $\Omega_{\rm Si} =  2.1 \times 10^{-7}$.   If we divide these parameters 
by the {\it solar} mass fractions of C (18.2\%) and Si (4.34\%), the photoionized \Lya-forest absorbers at low redshift
traced by \CIII, \CIV, \SiIII, and \SiIV\ would correspond to total metal mass densities $\Omega_Z = 1.9 \times 10^{-6}$ 
from carbon or $\Omega_Z = 4.8 \times 10^{-6}$ from silicon.  To make these two metallicity estimates consistent, we 
require Si/C to be enhanced by a factor of 3 relative to solar abundances.   Such enhancements of alpha-process 
elements, [O/Fe] $\approx+0.5$, have also been seen in metal-poor halo stars (Akerman \etal\ 2004)
and in low-metallicity damped \Lya\ absorbers (Pettini \etal\ 2008).  For example, in a survey of 20 metal-poor DLAs,  
Cooke \etal\ (2011) found a mean value [O/Fe] = $+0.39 \pm0.12$.   

We then re-calculate the metal fractions, enhancing the abundances by a factor of 3 for all alpha-process elements:   
even-$Z$ nuclei from atomic number $Z = 8$ to 20 (O through Ca).  The mass fractions become 7.95\% (C),  
5.70\% (Si), and 58.0\% (O), and the inferred metal abundances are $\Omega_Z =  4.3 \times 10^{-6}$ from carbon and 
$\Omega_Z = 3.7 \times 10^{-6}$ from silicon.   With the Si/C enhancement, we infer a consistent value 
$\Omega_Z = (4.0\pm0.5) \times 10^{-6}$ in the photoionized gas, or a mass density
$(5.4\pm 0.7) \times 10^5\;M_{\odot}\;{\rm Mpc}^{-3}$.  
We can perform a similar estimate of metal abundance from \OVI, the single observed ion stage of oxygen.   Numerous 
observations and models suggest that \OVI\ can be formed from both photoionization and collisional ionization.  As
shown in Figure 7, the fraction of \OVI\ from photoionization is expected to be small in the photoionized absorbers
at $\Delta_b = 200\pm50$.  In the hot gas, reflecting the fragility of  its Li-like ionization state, the maximum \OVI\ 
abundance fraction in collisional ionization equilibrium is $f_{\rm OVI} \approx 0.2$.  Cosmological N-body hydrodynamic 
simulations (Smith \etal\ 2011; Shull \etal\ 2012) of the low-redshift IGM, accounting for both photoionization and collisional
ionization, show that \OVI\  exists in multiphase gas, 
with inhomogeneous distributions of metallicity $(Z/Z_{\odot})$ and ionization fraction $f_{\rm OVI}$.  
In these calculations, the IGM-averaged product is $\langle (Z/Z_{\odot}) f_{\rm OVI} \rangle = 0.01$, a factor of 2 lower 
than the frequently assumed values of $Z/Z_{\odot} = 0.1$ and $f_{\rm OVI} = 0.2$.  Here, we assume an ionization 
fraction $f_{\rm OVI} = \Omega_{\rm OVI}  / \Omega_{\rm O} = 0.1$ and an oxygen-to-metals fraction 
$\Omega_{\rm O} /\Omega_Z = 0.58$, appropriate for all alpha-process elements enhanced by a factor of 3.  The observed
value in Table~2, $\Omega_{\rm OVI} = 38.6^{+4.8}_{-3.2}$ ($10^{-8} h_{70}^{-1}$) corresponds to 
$\Omega_{\rm O} = (3.9 \pm 0.5) \times 10^{-6}$ and total metal abundance $\Omega_Z = (6.7 \pm 0.8) \times 10^{-6}$.  
The metal density in the \OVI-traced gas is $\rho_Z = (9.1\pm0.6) \times 10^5~M_{\odot}~{\rm Mpc}^{-3}$.  

For the total IGM metal density, we combine the values for photoionized gas (traced by the C and Si ions) with the hot
gas (traced by \OVI).  The total metal density in both phases of the low-$z$ IGM is
$\rho_Z = (1.4\pm0.9) \times 10^6~M_{\odot}~{\rm Mpc}^{-3}$ or $\Omega_Z \approx 10^{-5}$.   As we discuss in
the next section, this corresponds to $\sim$10\% of the expected heavy elements produced by cosmic star formation,
integrated from $z \approx 8$ down to $z = 0$.  The resulting metal density, 
$(1.5 \pm 0.8) \times 10^7~M_{\odot}~{\rm Mpc}^{-3}$, is estimated from the integrated SFR density, 
$\rho_* = (6 \pm 2)\times 10^8~M_{\odot}~{\rm Mpc}^{-3}$, assuming a mean metal yield $y_m = 0.025\pm0.010$.

\subsection{Metal Production History}  

We now estimate the global metal-production, based on cosmological parameters and observations of the cosmic SFR 
history.  Several groups have integrated the metal-production density.   Pettini (1999) suggested that
$\rho_Z \approx 4.5 \times 10^6~M_{\odot}~{\rm Mpc}^{-3}$ of metals were produced by $z = 2.5$, with a metal-production 
yield $y_m = 1/42$ from  star formation between 11-13 Gyr.  In a follow-up analysis,  Pettini (2006) revised this to 
$3.4 \times 10^6~M_{\odot}~{\rm Mpc}^{-3}$ now assuming $y_m = 1/64$.  Bouch\'e \etal\ (2007) integrated 
the parameterized SFR history from Cole \etal\ (2001) with $y_m = 1/42$ to find $4\times10^6~M_{\odot}~{\rm Mpc}^{-3}$ 
down to $z = 2$ and $2.13 \times 10^7~M_{\odot}~{\rm Mpc}^{-3}$ down to $z = 0$.

In our current low-redshift IGM survey, we are interested in the star formation and metal production down to $z = 0$.  We 
adopt somewhat larger error bars on $\rho_*$ and $\rho_Z$ that reflect systematic uncertainties in measuring the SFR 
history and computing the metal yield.  These include a large (typically factor-of-five) dust correction to the SFR history
applied between $1 < z < 4$ (Bouwens \etal\ 2011).   Translating galaxy luminosity density into mass density also requires
assumptions about the stellar initial mass function (IMF), including its mass range, slope, and low-mass turnover. Similar
assumptions affect the metal yield (Madau \& Dickinson 2014) which can range from $y_m = 0.016-0.032$ for a Salpeter 
or Chabrier IMF counting the metals produced by massive stars between $10-40~M_{\odot}$.  Metal contributions from higher 
mass stars are cut off by core collapse into black holes. The range increases to $y_m = 0.023-0.048$ if the black hole cutoff 
is raised to $60~M_{\odot}$.   Studies of the dependence of metal production on cutoff (Brown \& Woosley 
2013) suggest a range $25\;M_{\odot} < M_{\rm BH} < 60\;M_{\odot}$.  Some of the uncertainty in metal production
is offset with the conversion from UV light to mass, since the same massive stars produce both UV photons and metals 
(Madau \& Shull 1996).  

Adding appropriate uncertainties in the SFR measurements and dust corrections, we adopt an integrated stellar mass 
density $\rho_* \approx (6 \pm 2) \times 10^8~M_{\odot}~{\rm Mpc}^{-3}$.  In their review, Madau \& Dickinson (2014) 
quote $\rho_* = 5.8 \times 10^8~M_{\odot}~{\rm Mpc}^{-3}$ from their model SFR history, in good agreement with the 
value $(6.0\pm1.0) \times 10^8~M_{\odot}~{\rm Mpc}^{-3}$ found by Gallazzi \etal\ (2008).    As a final check, we
integrated the SFR history (Bouwens \etal\ 2011) from $z = 8$ to the present.  Over the past 10 Gyr, 
the SFR density, expressed in units $M_{\odot}~{\rm yr}^{-1}~{\rm Mpc}^{-3}$, falls dramatically, from 
$\dot{\rho}_* \approx 0.1$ at $t = 10$~Gyr  ($z = 1.76$) to $\dot{\rho}_* \approx 0.01$ ($z \approx 0$).  The SFR density is 
well-fitted in look-back time ($t$) over the range $0 < t < 10~{\rm Gyr}$ by $\log \dot {\rho}_* = -2.0 + (t / 10~{\rm Gyr})$, 
equivalent to the exponential form:   
\begin{equation}
  \dot{\rho}_*  = (0.01~M_{\odot} ~ {\rm yr}^{-1} ~{\rm Mpc}^{-3} ) \,  \exp \, [ 2.3026 \, ( t / {\rm 10~Gyr})]     \; .
\end{equation} 
Integrated back to 10 Gyr  ($0 < z < 1.76$), this gives a total density of star formation,
${\rho}_*  = (0.01) (10^{10}~{\rm yr})  (9 / 2.3026) = 3.9 \times 10^8~M_{\odot}  ~{\rm Mpc}^{-3}$.
From $t = 10-11$~Gyr ($1.76 < z < 2.4$) the SFR density is near its peak value 
($0.1~M_{\odot} ~ {\rm yr}^{-1} ~{\rm Mpc}^{-3}$) and contributes $\sim1 \times 10^8~M_{\odot}  ~{\rm Mpc}^{-3}$.
The declining star formation from $z = 2.4$ to $z \approx 8$ (Bouwens \etal\ 2011) adds a similar metal density,
for an integrated total of $6 \times 10^8~M_{\odot}~{\rm Mpc}^{-3}$.    \\

Taking into account uncertainties in IMF and metal production, we adopt a metal yield $y_m = 0.025\pm0.010$ and
predict an integrated metal density $\rho_Z = (1.5 \pm 0.8) \times 10^7 M_{\odot}~{\rm Mpc}^{-3}$.  This corresponds to 
a fractional metal density $\Omega_Z  = (1.1 \pm 0.7) \times10^{-4}$ with combined uncertainties in integrated SFR and 
metal yield.   Some of these metals stay locked into stellar remnants, others remain within the galactic interstellar medium, 
and some are blown into the CGM and IGM.   Estimates of the metal fractions in these components range from 60-70\%
(Bouch\'e \etal\ 2007; Peeples \etal\ 2014).  For the CGM and IGM surveys, the primary issues are:  
(1) How many metals were expelled by galactic winds?   (2)  In what thermal phase and ionization state do they reside?  
(3) How many metals are undetected owing to their existence in higher ionization states?  \\

A considerable mass of metals resides in the halos of galaxies, in the CGM, and probably gas expelled to the IGM.
We estimate the galactic-halo contribution from the luminosity function of the Millennium Survey (Driver \etal\ 2005) 
with their Schechter-function parameters ($\phi^* = 0.0177 h^3~{\rm Mpc}^{-3}$, $\alpha = -1.13$).  We compute a 
galaxy space density $3.8\times10^{-3} \, h_{70}^3~{\rm Mpc}^{-3}$, converting from $h$ to $h_{70}$ and integrating 
between $0.5-1.5 L^*$ (effective bandwidth of $0.63L^*$).  We multiply by an estimated fraction, $f_{\rm SF} \approx 0.3$, 
of active star-forming galaxies and the mean mass of metals per halo, $2.65 \times 10^7~M_{\odot}$, found in \OVI\ 
absorbers from the COS-Halos study (Tumlinson \etal\ 2011, 2013) and scaled to total metals using the likely range of
oxygen mass fractions ($50\pm10$\%).   We obtain a metal density $\rho_Z \approx  3 \times 10^4~M_{\odot}~{\rm Mpc}^{-3}$ 
corresponding to $\Omega_Z \approx 2 \times 10^{-7}$.  This is only a small fraction of the cosmic metal production, but 
there are large uncertainties in these estimates.   The star-forming fraction $f_{\rm SF}$ depends on which stellar mass $M^*$ 
one chooses (Ilbert \etal\  2013; Baldry \etal\ 2012), while the integrated SFR, with its dust correction, is probably 
uncertain by a factor of 2 (Karim \etal\  2011; Madau \& Dickinson 2014).  \\

Our best estimate is that the low-$z$ IGM contains $10\pm5$\% of the cosmic metals, which agrees with estimates
for the metal abundance in various thermal phases of the IGM.    For example, we can combine the mean IGM metallicity, 
$Z_{\rm IGM}$ to the solar metallicity, $Z_{\odot} \approx 0.0153$ by mass (Caffau \etal\ 2011) with the cosmological 
baryon density,  $\Omega_b \approx 0.0461$, measured by microwave background experiments (Hinshaw \etal\ 2013).  
Our recent baryon census (Shull \etal\ 2012) found that the low-$z$  \Lya\ forest contains $28\pm11$\% of the baryons,  
with the shock-heated WHIM traced by \OVI\ and broad \Lya\ absorbers containing  $25\pm8$\%.  The expected metal 
densities in the \Lya\ forest and \OVI-traced WHIM would then be 
\begin{eqnarray}
     \Omega_{Z}^{(\rm Lya)}  & \approx &  0.28 \, \Omega_b \,  Z_{\rm IGM} 
            \approx (2.0 \times 10^{-6})  \left( \frac {Z_{\rm IGM}}{0.01 Z_{\odot}} \right)    \nonumber \\
     \Omega_{Z}^{(\rm WHIM)} & \approx &   0.25 \, \Omega_b \,  Z_{\rm IGM} 
            \approx (1.8 \times 10^{-6})  \left( \frac {Z_{\rm IGM}}{0.01 Z_{\odot}} \right) \; .
\end{eqnarray}
In our survey, the observed ion states (\CIII,  \CIV\, \SiIII, \SiIV) with ionization corrections suggested that 4\% of the cosmic 
metals reside in the photoionized IGM, with an additional 6\% in hotter gas traced  by \OVI.   The inferred metal densities were
$(5.4 \pm 0.7)\times10^5~M_{\odot}\,{\rm Mpc}^{-3}$ or $\Omega_Z = (4 \pm 0.5) \times 10^{-6}$ for the 
\Lya\ absorbers and
$(9.1\pm 0.6) \times10^5~M_{\odot}\,{\rm Mpc}^{-3}$ or $\Omega_Z = (6.7 \pm 0.5) \times 10^{-6}$ for the
\OVI-traced WHIM.   These densities correspond to mean metallicities of 2\% solar (\Lya\ forest) and 4\% solar
(WHIM),  both with substantial uncertainty from the ionization corrections.   One expects large variations in the 
IGM metallicity owing to incomplete metal transport and mixing.  \\

Finally, it is appropriate to ask whether the ions in this survey can be described, in either photoionization or collisional 
ionization equilibrium, with the alpha-enhanced abundance pattern as discussed above.  Most observations 
and cosmological simulations suggest that the low-$z$ IGM absorbers are multi-phase gas, with 
contributions from photoionization at $T \approx 10^4$~K (for \HI, \CIII, \CIV, \SiIII, \SiIV)  and hotter, shock-heated gas 
at $T \approx 10^{5.5\pm0.5}$~K containing much of the \OVI.   The oxygen ionization correction is less certain than 
those of C and Si; some of the \OVI\ may be photoionized and double-counted in the \Lya\ forest traced by C and Si ions.   
Below, we compare ratios of the {\it observed} ion states for the elements (O/C), (O/Si), (C/Si), (N/C), and (N/O) to their 
values assuming relative solar abundances:
\begin{eqnarray}
({\rm O/C)}  &=& \Omega_{\rm OVI} / (\Omega_{\rm CIII} + \Omega_{\rm CIV})  \approx  2.24^{+0.80}_{-0.40}    
                   \nonumber   \\
({\rm O/Si)} &=& \Omega_{\rm OVI} / (\Omega_{\rm SiIII} + \Omega_{\rm SiIV}) \approx  11.0^{+3.5}_{-1.9}      
                 \nonumber   \\
({\rm C/Si)} &=& (\Omega_{\rm CIII} + \Omega_{\rm CIV}) / (\Omega_{\rm SiIII} + \Omega_{\rm SiIV}) 
                 \approx 4.9^{+2.2}_{-1.1}   \nonumber   \\
({\rm N/C)}  &=& \Omega_{\rm NV} / (\Omega_{\rm CIII} + \Omega_{\rm CIV}) \approx  0.11^{+0.05}_{-0.03}  
                   \nonumber   \\
({\rm N/O)} &=& \Omega_{\rm NV}  / \Omega_{\rm OVI} \approx  0.049^{+0.017}_{-0.011}  
\end{eqnarray} 
For reference, the solar mass-abundance ratios can be derived from the abundances by number (Caffau \etal\ 2011):
$\rho_{\rm O} / \rho_{\rm C} = (16/12) (n_{\rm O} / n_{\rm C}) =  2.43$,   
$\rho_{\rm O} / \rho_{\rm Si} = (16/28) (n_{\rm O} / n_{\rm Si}) = 8.65$,
$ \rho_{\rm C} / \rho_{\rm Si} = (12/28) (n_{\rm C} / n_{\rm Si}) =  4.18$,   
$\rho_{\rm N} / \rho_{\rm C} = (14/12) (n_{\rm N} / n_{\rm C}) =  0.25$, and 
$\rho_{\rm N} / \rho_{\rm O} = (14/16) (n_{\rm N} / n_{\rm O}) =  0.12$.  
From photoionization modeling of the C and Si ions\footnote{Because the ionization corrections for C and Si 
respond in different ways to EUV and X-ray photoionization (see Appendix B), the Si/C abundance ratio inferred 
from just the two observed ion states, \CIII\ and \CIV\ compared to \SiIII\ and \SiIV, is not an accurate measure 
of the total abundances.} we only obtain a consistent metallicity if Si/C is enhanced by a factor of 3 over solar values.  
The (O/C) and (C/Si) ratios are comparable to the solar abundances (Asplund \etal\ 2009; Caffau \etal\ 2011), while 
the (O/Si) ratio is somewhat larger.  As often is the case, N is underabundant relative to both C and O, although with 
just a single measured ion state (\NV\ and \OVI) these ratios are not necessarily accurate indicators of total abundances.   
However, the observed mass-density ratio, $\Omega_{\rm NV} / \Omega_{\rm OVI} \approx 0.049$ may indicate that (N/O)
$\approx$ 40\% of the solar ratio.  From an analysis (Pettini \& Cooke 2012) of the nitrogen metallicity dependence 
on primary and secondary nucleosynthesis, this ratio suggests that the sources of this gas likely had metallicity well 
above the metal-poor floor at $\log ({\rm N/O}) = -2.3$.

\section{CONCLUSIONS AND DISCUSSION} 

A fundamental uncertainty in any  ``missing metals" survey is set by estimates of the total metal production
rates from the integrated SFR history and metal yields.   Our UV survey finds that the low-$z$ IGM contains 
$10\pm5$\% of the metal density $\rho_Z = (1.5 \pm 0.8) \times 10^7\;  M_{\odot}~{\rm Mpc}^{-3}$ integrated
to $z = 0$.  The bulk of the metals are likely contained in stars and ISM within galaxies (Bouch\`e \etal\ 2007) and 
in hot gas in galactic halos and the CGM (Peeples \etal\ 2014).  The efficiency of metal transport is still uncertain, 
as recently injected metals may not extend from their galactic sources, given the outflow speeds and finite lifetimes 
of the flows (Oppenheimer \& Dav\'e 2008).  Thus, it is no surprise that only 10\% of the metals have made it into
the IGM, with many more located in galaxy halos and the CGM (Tumlinson \etal\ 2013; Stocke \etal\ 2013). \\

Our \HST/COS survey of QSO metal-line absorbers provides the largest current sample of metallicity in the low-redshift 
IGM.  Of particular importance are abundances of adjacent ion stages, whose ratios \CIII/\CIV\ and \SiIII/\SiIV\  improve
the ionization corrections.   Because the production of high ions of C and Si have different dependences on EUV and 
X-ray photons, their individual ionization corrections constrain the ionizing background and baryon overdensity.  We
correct the measured ion densities of the adjacent ion states, 
$(\Omega_{\rm CIII} + \Omega_{\rm CIV})$ and $(\Omega_{\rm SiIII} + \Omega_{\rm SiIV})$, by factors 
$\Omega_{\rm C}  / \left[  \Omega_{\rm CIII} +  \Omega_{\rm CIV} \right] =  2.0^{+1.0}_{-0.5}$ and 
$\Omega_{\rm Si} / \left[  \Omega_{\rm SiIII} +  \Omega_{\rm SiIV} \right] =  6^{+4}_{-3}$ to derive elemental mass densities 
$\Omega_{\rm C} =  3.4 \times 10^{-7}$ and $\Omega_{\rm Si} =  2.1 \times 10^{-7}$.  Consistency of the C and Si
metallicities requires that Si/C be enhanced by a factor 3 over solar abundances.    Presumably, other alpha-process 
elements (even-$Z$ nuclei from O through Ca) are also enhanced.  The inferred Si/C enhancement in the low-$z$ IGM 
suggests that these metals were injected at the peak of star formation ($z \approx 1-3$) by young stellar populations whose
massive stars provided alpha-process nucleosynthesis. \\

The total metal abundance in the low-$z$, photoionized  \Lya\ forest absorbers is $\Omega_Z = (4.0 \pm 0.5) \times 10^{-6}$, 
consistent with the individual values $\Omega_Z =  4.3 \times 10^{-6}$ inferred from carbon and $\Omega_Z = 3.7 \times 10^{-6}$
from silicon.  A similar exercise for  \OVI\ gives  $\Omega_{\rm O} = 3.9 \times 10^{-6}$ and $\Omega_Z = 6.7 \times 10^{-6}$.  
Because \OVI\ likely traces a different thermal phase than the photoionized C and Si ions, these two densities are additive, 
yielding a total IGM metal density of $\Omega_Z \approx 10^{-5}$.  This density is $\sim$10\% of the metals produced by 
cosmic star formation.  Additional metals reside in the halos and CGM of galaxies as probed in absorption lines of \OVI\  
(Tumlinson \etal\ 2013), \CIV\ (Borthakur \etal\ 2013), and \CaII\ (Zhu \& M\'enard 2013).   Quantitative estimates of the 
total metal content in these reservoirs will require UV and X-ray absorption-line measurements that cover the range of
ionization states in this multiphase gas (Stocke \etal\ 2013; Werk \etal\ 2014) located 50-300~kpc from galaxies, both within
and beyond their virial radii (Shull 2014).  

\newpage

\noindent
The primary conclusions of our low-redshift IGM metal survey are as follows:
\begin{enumerate}

\item  For six metal ions in the IGM surveyed by HST/COS spectra (Danforth \etal\ 2014) at  
$\langle z \rangle \approx 0.14$, we find cosmic mass densities 
$\Omega_{\rm ion} \equiv \rho_{\rm ion} / \rho_{\rm cr}$ (in units $10^{-8} \, h_{70}^{-1}$) of
$\Omega_{\rm CIV} = 10.1^{+5.6}_{-2.4}$, $\Omega_{\rm CIII} = 7.1^{+1.9}_{-1.2}$, 
$\Omega_{\rm OVI} = 38.6^{+4.8}_{-3.2}$, $\Omega_{\rm SiIV} = 2.1^{+1.0}_{-0.5}$, 
$\Omega_{\rm SiIII} = 1.4^{+0.3}_{-0.2}$, and $\Omega_{\rm NV} = 1.9^{+0.6}_{-0.4}$ integrated over
column density ranges specified in Table 2.  

\item  Our survey includes key intermediate ion stages (\CIII\ and \SiIII) as well as \CIV\ and \SiIV.  The 
mean mass-abundance ratios of the survey, \CIII/\CIV\ $\approx 0.70^{+0.43}_{-0.20}$ 
and \SiIII/\SiIV\ $\approx 0.67^{+0.35}_{-0.19}$, are consistent with a metagalactic ionizing background 
with photoionization parameter $\log U = -1.5 \pm 0.4$.   A suite of photoionization models yields correction factors
for higher ion stages and requires that Si/C be enhanced by a factor of 3 above solar abundances.
 
\item Applying photoionization corrections to (\CIII\ + \CIV) and (\SiIII\ + \SiIV), we estimate a metal density 
of $\Omega_Z \approx (4.0 \pm 0.5) \times 10^{-6}$ or  4\% of the predicted metal production.  For hot IGM traced
by \OVI, we find $\Omega_Z \approx (6.7\pm0.8)  \times 10^{-6}$ or 6\% of the metal production.  Combining the 
two reservoirs, we estimate that the low-$z$ IGM contains $10\pm5$\% of the total metal production, 
$(1.5 \pm 0.8) \times 10^7~M_{\odot}~{\rm Mpc}^{-3}$, predicted from integrated star formation, 
$(6 \pm 2)\times 10^8~{M_{\odot}~\rm Mpc}^{-3}$ with yields  $y_m \approx 0.025 \pm 0.010$. 

\item  For several ionization backgrounds (at $z \approx 0$), our photoionization models of the metal absorbers 
are consistent with a soft ionizing background above the \HeII\ edge with ionizing intensity 
$I_0 = (3\pm1) \times 10^{-23} \; {\rm erg~cm}^{-2}~{\rm s}^{-1}~{\rm Hz}^{-1}~{\rm sr}^{-1}$ at the hydrogen Lyman limit.
These backgrounds correspond to one-sided ionizing flux $\Phi_0 \approx 10^4$ cm$^{-2}$~s$^{-1}$ and 
hydrogen ionization rate $\Gamma_{\rm H} \approx (8 \pm 2) \times 10^{-14}$ s$^{-1}$.  These values need additional 
observational constraints, but they are consistent with previous theoretical calculations (Haardt \& Madau 2001; 
Shull \etal\ 1999) and inferences from the \Lya\ column-density distribution (Kollmeier \etal\ 2014; Egan \etal\ 2014).  

\item The photoionization parameter $\log U = -1.5\pm0.4$ with baryon overdensities $\Delta_b \approx 200\pm50$
and Si/C =  3 times solar abundances.   This overdensity corresponds to $n_H \approx 10^{-4.25\pm0.10}$ cm$^{-3}$ 
at $\langle z \rangle = 0.14$, a hydrogen gas density typical of that in extended galactic halo gas.  

\item Given the uncertainties in integrated SFR and metal production and the demonstrated existence of reservoirs of 
metals (galactic stars, galaxy halos, CGM, and hotter gas), there is probably no compelling reason to require a
missing metals problem.  Further UV and X-ray measurements of these galactic and CGM reservoirs and their
ionization conditions will be needed to quantify this metal census.  

\end{enumerate}

Looking toward future improvements in these metal surveys, we note the uncertainties in the metal-ion column 
density distributions at the high end:  $\log N  > 15.0$ (\CIV\ and \OVI),  $\log N > 13.5$ (\SiIII),  $\log N > 14.0$ (\SiIV), 
and $\log N > 14.5$ (\CIII).  From power-law fits to these CDDFs, we see that better characterization would be most useful
for \SiIV,  \NV, and \SiIII.   The Si abundances are less secure than those of C, owing to a smaller number of \SiIV\ 
absorbers and consistent with the scatter in low-$z$ measurements of $\Omega_{\rm SiIV}$ (Danforth \& Shull 2008; 
Cooksey \etal\ 2011; Tilton \etal\ 2012; Danforth \etal\ 2014).  The total redshift pathlengths in the medium-resolution 
COS bands (G130M and G160M) from 1135-1796~\AA\ vary considerably  by ion:  $\Delta z \approx 20$ (\HI), 18 (\SiIII), 
13 (\SiIV), 17 (\NV), 13 (\OVI), 10 (\CIII), and 8 (\CIV).  The lines of \CIV\ and \SiIV\ redshift out of the G160M band at 
$z > 0.16$ and $z > 0.29$ respectively.
Longer wavelength \HST\ surveys would require considerably more observing time with less efficient gratings.  
Additional ionization states would also help to characterize the ionization modeling.  In the UV, surveys of the lower 
ions (\CII\ and \SiII) would complement our measurements of intermediate ion states (\CIII, \CIV, \SiIII, \SiIV).  To better
constrain the uncertain high-energy background radiation and inner-shell ionization,  it would be helpful to observe the 
expected higher ion states in X-ray absorbers (\CV, \CVI,  \OVII, \OVIII, \SiV, \SiVI, \SiVII).

\acknowledgments

\noindent
We thank George Becker, Alec Boksenberg, Bob Carswell, Mark Giroux, Max Pettini, and Joop Schaye for helpful 
discussions and comments on metal-line measurements in the IGM and ionizing backgrounds.  
This research was supported by the STScI COS grant (NNX08-AC14G) at the University of Colorado Boulder.
JMS thanks the Institute of Astronomy at Cambridge University for their stimulating scientific environment
and support through the Sackler Visitor Program.

\newpage


\appendix 


\section{Appendix~A:  Analytic Estimates of $\Omega_{\rm ion}$} 

Equation (1) gives an expression for the mass density of an ion, involving an integral over the column density
distribution function (CDDF), denoted by $f(N,z) \equiv (\partial ^2 {\cal N} / \partial N \, \partial z)$.   In our IGM
survey (Danforth \etal\  2014), we fitted the cumulative distribution of column densities ($N$) to a power law,
\begin{equation}
     \frac { d {\cal N} (>N) } { dz }  = C_0   \left( \frac {N} { N_0} \right) ^{-(\beta-1)}    \; \;  , 
\end{equation}
where $C_0$ is the normalization at fiducial column density  $N_0 = 10^{14}~{\rm cm}^{-2}$.  For several ions 
(\OVI\ and \CIV) we fitted piecewise continuous power laws, for strong and weak absorbers.  We obtain the CCDF 
by differentiating the cumulative distribution with respect 
to $N$, 
\begin{equation}
   f(N,z) = \frac {C_0 (\beta - 1)}{N_0}   \left( \frac {N}{N_0} \right) ^{-\beta}   \; .
\end{equation} 
With $x = N/N_0$ and $N_0 = 10^{14}~{\rm cm}^{-2}$, the fractional ion density becomes
\begin{eqnarray}
     \Omega_{\rm ion} &=& (1.365 \times 10^{23}~h_{70}^{-1} \,{\rm cm}^2)  \left( \frac {m_{\rm ion}} {{\rm amu}} \right)
           \int_{\rm N_{\rm min}}^{\rm N_{\rm max}}  \frac {C_0 (\beta-1) }{N_0} \left( \frac {N}{N_0} \right) ^{-\beta}  \,  N \,  dN  \nonumber \\
      &=& (1.365 \times 10^{-9}~h_{70}^{-1})  \, C_0 (\beta-1)   \left( \frac {m_{\rm ion}} {{\rm amu}} \right) 
            \int_{\rm x_{\rm min}}^{\rm x_{\rm max}}  x^{-(\beta-1)} \, dx   \; ,
\end{eqnarray}
an expression that can be evaluated for each of the segments of the CCDF fits.  From this integral, we see that distributions 
with $\beta < 2$ are dominated by high-$N$ absorbers (near $N_{\rm max}$) while expressions with $\beta > 2$ are dominated 
by low-$N$ absorbers (near $N_{\rm min}$).  
\begin{eqnarray}
   \Omega_{\rm ion} &=& (1.365 \times 10^{-9}~h_{70}^{-1})  \, C_0  \left( \frac  {\beta-1} {\beta-2}  \right) 
           \left( \frac {m_{\rm ion}} {{\rm amu}} \right) \left[ x_{\rm min}^{-(\beta-2)} - x_{\rm max}^{-(\beta-2)} \right] \; \; 
               ({\rm for}~\beta > 2)  \\
   \Omega_{\rm ion} &=& (1.365 \times 10^{-9}~h_{70}^{-1})  \, C_0  \left( \frac  {\beta-1}{2 - \beta}  \right)   
           \left( \frac {m_{\rm ion}} {{\rm amu}} \right) \left[ x_{\rm max}^{(2-\beta)} - x_{\rm min}^{(2-\beta)} \right] \; \; 
               ({\rm for}~ \beta <  2)    \\
    \Omega_{\rm ion} &=& (1.365 \times 10^{-9}~h_{70}^{-1})  \, C_0  
           \left( \frac {m_{\rm ion}} {{\rm amu}} \right)  \ln \left( \frac {x_{\rm max}} {x_{\rm min}}  \right)  \; \;  ({\rm for}~\beta = 2)  
\end{eqnarray}        
Table 3 gives the fitting parameters ($C_0$, $\beta$) for the metal six ions in our survey and the resulting values of 
$\Omega_{\rm ion}$, which can be compared with the discrete numerical integrals over logarithmic bins, $\Delta \log N  = 0.2$,  
given in Table 2.

\newpage


\section{Appendix~B:   Photoionization Rates and Ionizing Fluxes}  

In this Appendix  we discuss analytic approximation to the equilibrium ionization ratios of high ions of C and Si.  
In the approximation that each ionization stage of an element is coupled only to those immediately above 
and below, the ionization fractions, $f_i$, can be derived by solving pairwise along the ``rungs of the ladder".    
Consider two adjacent ion stages, \SiIV\ and \SiV, in photoionization equilibrium with abundance ratio
$n_{\rm SiV} /  n_{\rm SiIV} \approx (\Gamma_{\rm SiIV} /  n_e \alpha_{\rm SiIV})$.  Here, 
$\Gamma_{\rm SiIV}$ (s$^{-1}$) is the photoionization rate of \SiIV\  and $\alpha_{\rm SiIV}$ (cm$^3$~s$^{-1}$) 
is the recombination rate coefficient from \SiV\ to \SiIV.  For an isotropic radiation field with specific intensity of 
power-law form, $I_{\nu} = I_i (\nu / \nu_T)^{-\alpha}$, at energies $h \nu \geq h \nu_T$ above the
ionization threshold ($E_T = h \nu_T$) of \SiIV, and for photoionization cross sections fitted to power-law form, 
$\sigma_{\nu} = \sigma_T (\nu / \nu_T)^{-\beta}$, one can derive the photoionization rate, 
\begin{equation}  
   \Gamma_{\rm SiIV} =  4 \pi  \int_{\nu_0}^{\infty} \frac { I_{\nu} \sigma_{\nu} } {h \nu }  \; d \nu 
       = \frac {4 \pi I_i \sigma_i} {h (\alpha + \beta)} 
       = (1.90 \times 10^{-14}~{\rm s}^{-1})  \left[ \frac {I_{-23} \, \xi_{\rm SiIV} }
             { (\alpha + \beta) } \right] \left( \frac  { \sigma_T}{ {\rm Mb}} \right)  \; .
\end{equation} 
The photoionization cross section is scaled to a threshold value of 1 Mb $(10^{-18}~{\rm cm}^2)$, where a fit 
to the \SiIV\ tabulation by Verner \etal\ (1996)  gives $\sigma_T =  0.314$~Mb and $\beta \approx 1.15$
between $E_T$ and $1.4 E_T$.  The intensity at the \SiIV\ threshold 
($E_T = 45.14$~eV) is written $I_i = J_0 \, \xi_{\rm SiIV}$, reduced by a factor $\xi_{\rm SiIV}$ from 
its value $I_0 = (10^{-23}~{\rm erg~cm}^{-2}~{\rm s}^{-1}~{\rm Hz}^{-1}~{\rm sr}^{-1}) J_{-23}$ at the hydrogen 
Lyman limit ($h \nu_0 = 13.60$~eV).  For a power-law radiation field, 
$\xi_{\rm SiIV} = (\nu_{\rm SiIV} / \nu_0)^{-\alpha} \approx 0.186$ for $\alpha = 1.4$, and 
$\Gamma_{\rm SiIV} = (4.35 \times 10^{-16}$~s$^{-1}) I_{-23} (\xi_{\rm SiIV} / 0.186)$.
The recombination rate coefficient at temperature  $T = (10^4~{\rm K})T_4$ is approximated (Shull \& Van 
Steenberg 1982) as $\alpha_{\rm SiIV} \approx (5.5 \times 10^{-12}$ cm$^{3}$~s$^{-1}) T_4^{-0.821}$.  
The electron density in the IGM absorber
$n_e = 1.167 \bar{n}_H \Delta_b = (2.22 \times 10^{-7}~{\rm cm}^{-3}) \Delta_b$,
for overdensity $\Delta_b = 100 \Delta_{100}$ and fully ionized helium with abundance 
$y = n_{\rm He}/n_{\rm H} = 0.0833$. Thus, for $(\alpha + \beta) = 2.55$, we have
\begin{equation}
    \frac {n_{\rm SiV}} {n_{\rm SiIV}} =  \frac  { \Gamma_{\rm SiIV}} { n_e \alpha_{\rm SiIV}} 
      \approx   (3.6) I_{-23} \,  \Delta_{100}^{-1}  \;  T_4^{0.821} \;  (\xi_{\rm SiIV}/0.186) \; .
\end{equation}  

One can perform the same exercise for the \SiV\ and \SiVI\ pair, where a fit to \SiV\ photoionization 
cross sections (Verner \etal\ 1996) gives $\sigma_T = 2.61$~Mb at threshold, $E_T = 166.77$~eV,
rising to 3.2-3.4~Mb between 170-200~eV, and then declining from 200-300 eV with
$\sigma_{\nu} \propto \nu^{-2.1}$.  The recombination rate coefficient 
$\alpha_{\rm SiV} = (1.2 \times 10^{-11}$ cm$^3$~s$^{-1}) T_4^{-0.735}$ and the \SiV\ 
flux-reduction factor is $\xi_{\rm SiV} = (166.77/13.60)^{-1.4} \approx 0.030$.  We estimate that
$\Gamma_{\rm SiV} \approx (5.6 \times 10^{-16}~{\rm s}^{-1})  (\xi_{\rm SiV}/0.030)$ and 
\begin{equation}
    \frac {n_{\rm SiVI}} {n_{\rm SiV}} = \frac  { \Gamma_{\rm SiV}} { n_e \alpha_{\rm SiV}} 
       \approx (2.1)  I_{-23} \, \Delta_{100}^{-1} \; T_4^{0.735} \;  (\xi_{\rm SiV}/0.030)  \; .
 \end{equation}
Evidently,  \SiV\ and \SiVI\ are likely to have significant ionization fractions for $\Delta_b \leq 100$, 
if the ionizing radiation field falls off as a power law, $(\nu / \nu_0)^{-\alpha}$ at $h \nu \gg 1$~ryd 
above the \SiIV\ and \SiV\ edges.  Inner-shell ionization by X-rays above the K-edges,
1910~eV, 1930~eV, and 1950~eV for \SiIII, \SiIV, and \SiV, respectively, followed by Auger electron 
emission (Weisheit 1974; Donahue \& Shull 1991) will enhance the \SiV\ and \SiVI\ abundances.  
Following K-shell ionization, \SiIII\ will jump to \SiVI\ or \SiVII\ with 2 or 3 Auger electrons released depending 
on whether the inner-shell (1s) vacancy is filled by a {\it 2s} or {\it 2p} electron.  The difference arises because 
of valence (3s) electrons, which can be released when a {\it 2p} electron drops into a {\it 2s} vacancy.  
Similarly, a K-shell ionization of \SiIV\  will release either 1 or 2 Auger electrons, producing \SiVI\ or \SiVII.  

Carbon, with fewer electrons, offers a somewhat different response to K-shell ionization, with threshold 
energies at 296 eV, 317 eV, and 347 eV for \CII, \CIII, and \CIV, respectively.  With three electrons, \CIV\ 
({\it 1s$^2$\,2s}) will release no Auger electrons to fill the K-shell vacancy, but \CIII\  ({\it 1s$^2$\,2s$^2$}) 
will release a single {\it 2s} electron through the Auger process.  
Thus, inner-shell ionization of \CIII\ and \CIV\ will both produce \CV.   For \CIV\ valence-shell ({\it 2s}) 
photoionization, $\sigma_T = 0.656$~Mb  at $E_T = 64.49$~eV, and $\beta \approx 1.7$ between $E_T$ 
and $1.4 E_T$ from the tabulated cross sections (Verner \etal\ 1996).  The radiative recombination rate 
coefficient $\alpha_{\rm CIV} \approx (7.5 \times 10^{-12}~{\rm cm}^3~{\rm s}^{-1}) T_4^{-0.817}$.  
Thus, for a $\nu^{-1.4}$ ($\alpha = 1.4$) radiation field, a flux-reduction factor 
$\xi_{\rm CIV} = (64.49/13.60)^{-1.4} \approx 0.113$, and $(\alpha + \beta) = 3.5$, the photoionization rate 
$\Gamma_{\rm CIV} \approx (4.54 \times 10^{-16}~{\rm s}^{-1})  I_{-23} \;  ( \xi_{\rm CIV} / 0.113)$ and we have
\begin{equation}
    \frac {n_{\rm CV}} {n_{\rm CIV}} =  \frac  { \Gamma_{\rm CV}} { n_e \alpha_{\rm CIV}} 
      \approx   (2.7) I_{-23} \,  \Delta_{100}^{-1}  \;  T_4^{0.817} \;  (\xi_{\rm CIV}/0.113) \; .
\end{equation}



\begin{deluxetable}{lclccr}
\tabletypesize{\footnotesize}
\tablecaption{\bf Absorption Line Data\tablenotemark{a}  }  
\tablecolumns{6}
\tablewidth{0pt}
\tablehead{   \colhead{Ion}    &    \colhead{$\lambda_0$~(\AA) }   &  \colhead{$f$ }  & $z$-range &
      \colhead{$\log N_{\rm min}$\tablenotemark{b} } &  \colhead{ $\tau_0$\tablenotemark{b} }  \\  
       &    &     & $z_{\rm min} - z_{\rm max}$  &  ($N$ in cm$^{-2}$)   &       
     }
   
\startdata
  C~III       &  977.020     &  0.757   &  0.16 - 0.84   &   12.67     &  0.443         \\
  C~IV      & 1548.204    &  0.176   &  0.00 - 0.16   &   12.87     &  0.176          \\
  N~V       & 1238.821    &  0.116   &  0.00 - 0.45   &   13.15     &  0.116           \\
 O~VI       & 1031.926    &  0.133   &  0.10 - 0.74   &  13.38      &  0.0819       \\
 Si~III       & 1206.500    &  1.63      & 0.00 - 0.49   &   12.15     &  1.18            \\
 Si~IV      & 1393.760     &  0.513   & 0.00 - 0.29   &   12.53    &  0.428          \\
\enddata 

\tablenotetext{a}{Columns (1)--(4) list observed ion, rest wavelength ($\lambda_0$), and absorption 
oscillator strength ($f$) from Morton (2003), and redshift range observable between 1135-1796~\AA\ 
in these primary transitions with the COS gratings (G130M and G160M). }

\tablenotetext{b}{$N_{\rm min}$ is the minimum column density corresponding to the 30~m\AA\ 
equivalent  width used as a lower cut-off for the survey.  Possible line saturation is gauged by the 
optical depth at line center, $\tau_0 = (5.99 \times 10^{-4} ) (N_{13} \,  f \lambda / b_{25})$, where 
$N_{13}$ is the column density in units of $10^{13}$~cm$^{-2}$, $\lambda$ is in \AA, and $b_{25}$ 
is the Doppler parameter in units of 25~\kms.  }

\end{deluxetable}



\begin{deluxetable}{lrccc}
\tabletypesize{\footnotesize}
\tablecaption{\bf Summary of Metal-Ion Densities\tablenotemark{a} }  
\tablecolumns{5}
\tablewidth{0pt}
\tablehead{   \colhead{Ion\tablenotemark{a} }    &    \colhead{$N_{\rm abs}$\tablenotemark{a} }   &  
   \colhead{Range in ($\log N$)\tablenotemark{a} }  &  
    \colhead{$\Omega_{\rm ion}{\rm (STIS)}$\tablenotemark{a}  }  &  
    \colhead{$\Omega_{\rm ion}{\rm (COS)}$\tablenotemark{a}  } \\  
   &   &   ($N$ in cm$^{-2}$)   &   (in $10^{-8}~h_{70}^{-1}$)  & (in $10^{-8}~h_{70}^{-1}$)  
     }
   
\startdata
  C~III       &   77    &  12.67--14.19    & $4.6^{+1.8}_{-0.8}$      &   $7.1^{+1.9}_{-1.2}$     \\ 
  C~IV      &  49     &  12.87--14.87    & $8.1^{+4.6}_{-1.7}$       &  $10.1 ^{+5.6}_{-2.4}$    \\
  N~V       &  37     &  13.15--14.01    & $0.9^{+1.3}_{-0.5}$       &  $1.9^{+0.6}_{-0.4}$    \\
 O~VI       & 212    &  13.38--14.83    & $33.9^{+9.3}_{-4.1}$    &   $38.6^{+4.8}_{-3.2}$    \\
 Si~III       &   87    &  12.15--13.44    & $1.1^{+0.8}_{-0.3}$       & $1.4^{+0.3}_{-0.2}$    \\
 Si~IV      &   31    &  12.53--13.94    & $4.5^{+3.0}_{-1.2}$       &  $2.1^{+1.0}_{-0.5}$    \\
\enddata

\tablenotetext{a}{Columns (1)--(3) list  observed ions, number of absorbers in COS survey in stated range of 
ion column densities ($\log N$).   Minimum  column density ($N_{\rm min}$) corresponds to 30~m\AA\ 
equivalent width in primary diagnostic line (Table 1). Maximum ($N_{\rm max}$) corresponds to last measured
absorber at which survey statistics are reliable.  Columns (4) and (5) list IGM mass density (at low $z$) of six 
ions, quoted as fractional contribution ($\Omega_{\rm ion} = \rho_{\rm ion} / \rho_{\rm cr}$) to closure density.  
These values are derived from HST/STIS survey (Tilton \etal\ 2012, revised as noted in text) and HST/COS survey 
(Danforth \etal\ 2014). 
    }

\end{deluxetable}



\begin{deluxetable}{lclcccccc}
\tabletypesize{\footnotesize}
\tablecaption{\bf Column-Density Distribution Fits\tablenotemark{a}  }  
\tablecolumns{9}
\tablewidth{0pt}
\tablehead{   \colhead{Ion}    &   \colhead{$C_0$(weak) }     &   \colhead{$\beta$(weak) }  &  \colhead{$C_0$(strong) } &  
                        \colhead{$\beta$(strong) }  &    \colhead{$\log N_b$ } &  $\log N_{\rm min}$ & $\log N_{\rm max}$ & 
                        \colhead{ $\Omega_{\rm ion}$\tablenotemark{b} }  \\
       &    &    &   &   &   &    & & ($10^{-8}\,h_{70}^{-1}$)          
     }
   
\startdata
  C~IV      & $2.7\pm1.6$   & $1.5\pm0.2$   &  $1.6\pm0.3$    &  $2.0\pm0.2$   &  13.546  & 12.87 & 15.0   & 10.2    \\
  Si~IV     & $0.2\pm0.1$   & $1.8\pm0.1$    &     \dots              &      \dots             &   \dots     &  12.53 & 15.0  & 3.2       \\
  O~VI      & $9.6\pm0.9$   & $1.6\pm0.1$   &  $11\pm2$        &   $3.9\pm0.6$  &  14.026  &  13.38 & 15.0  & 46.4     \\
  N~V       &  $0.4\pm0.1$  & $2.1\pm0.2$   &   \dots                 &   \dots                &   \dots     &  13.15 & 15.0  &  3.5    \\
\enddata 

\tablenotetext{a}{Power-law fits to the cumulative column density distribution, $d{\cal N}(>N)/dz$, of the form 
$C_0 (N/N_0)^{-\beta}$, with fiducial column $N_0 = 10^{14}$ cm$^{-2}$.   Ions \SiIV\ and \NV\ are fitted with a 
single power law, while \CIV\ and \OVI\ are fitted with a double power law (for weak and strong absorbers)
matched at a break column density $\log N_b$.   }

\tablenotetext{b}{Value of ion density parameter, $\Omega_{\rm ion}$, integrating  power-law distribution
 from $N_{\rm min}$ to $N_{\rm max}$, using Equations (A4), (A5), or (A6)  in Appendix A.   }

\end{deluxetable}


\clearpage


\begin{figure} 
 \includegraphics[scale=0.9]{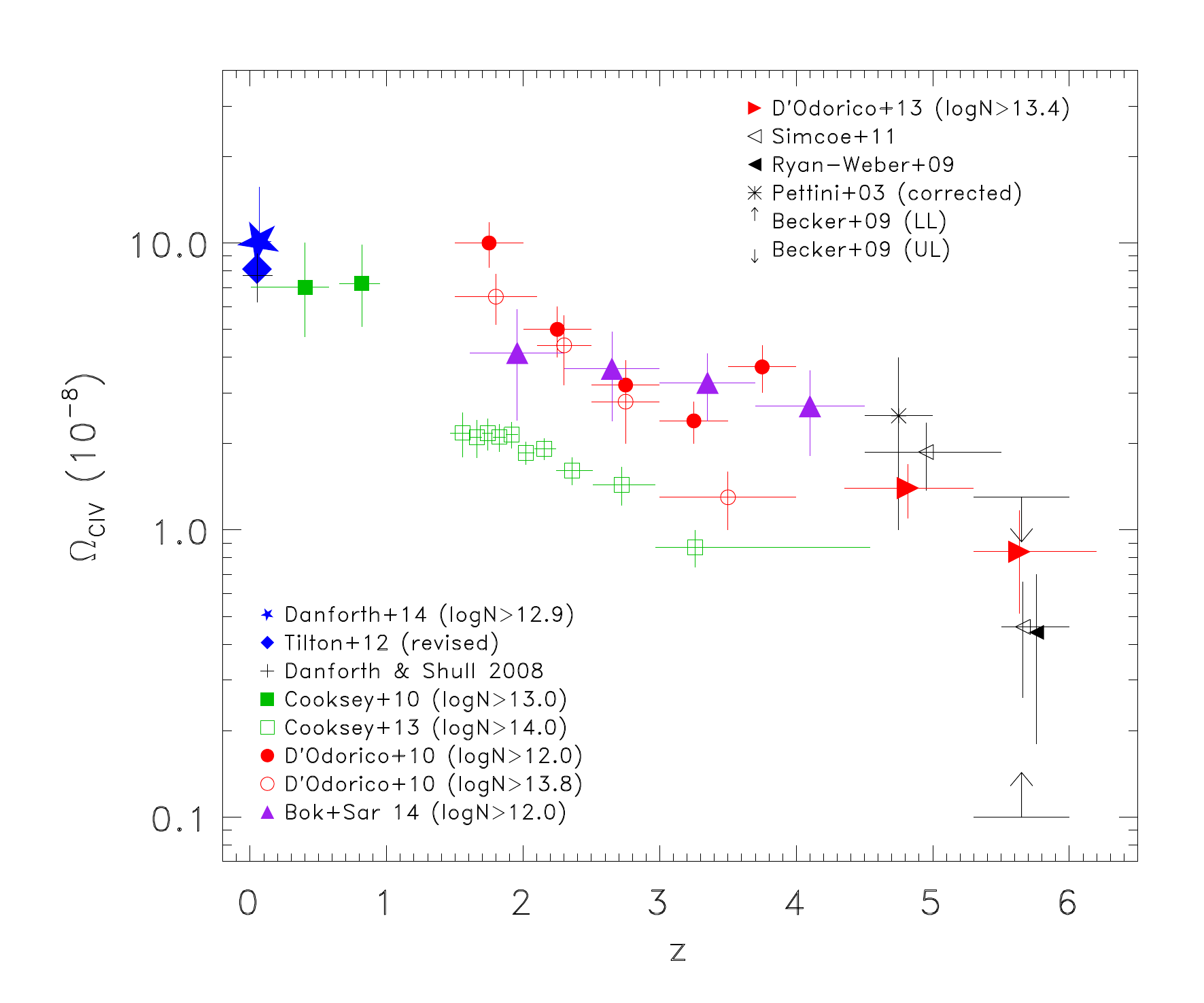}
\caption{\small Redshift evolution of \CIV\ mass density, with $\Omega_{\rm CIV}$ expressed relative to closure density
in units of $10^{-8} \, h_{70}^{-1}$.   Our HST/COS low-redshift measurement is shown together with values from past 
surveys.   Lower envelope includes a survey at $1.5 < z < 4$ (Cooksey \etal\ 2013) that measures strong absorbers, 
while upper envelope includes weaker absorbers, with column densities in various ranges, from $13.4 < \log N_{\rm CIV} < 15$ 
(D'Odorico \etal\ 2013), from 12-15 (D'Odorico \etal\  2010), 12.0-14.5 (Boksenberg \etal\ 2003), and 12.9-14.9 (Danforth
 \etal\  2014).   Taking into account corrections (see Appendix A) for these different N-ranges, the \CIV\ abundance 
 has evidently increased by a factor of 10 since $z \approx 5$.   }
\end{figure}



\begin{figure}
\includegraphics[scale=0.9]{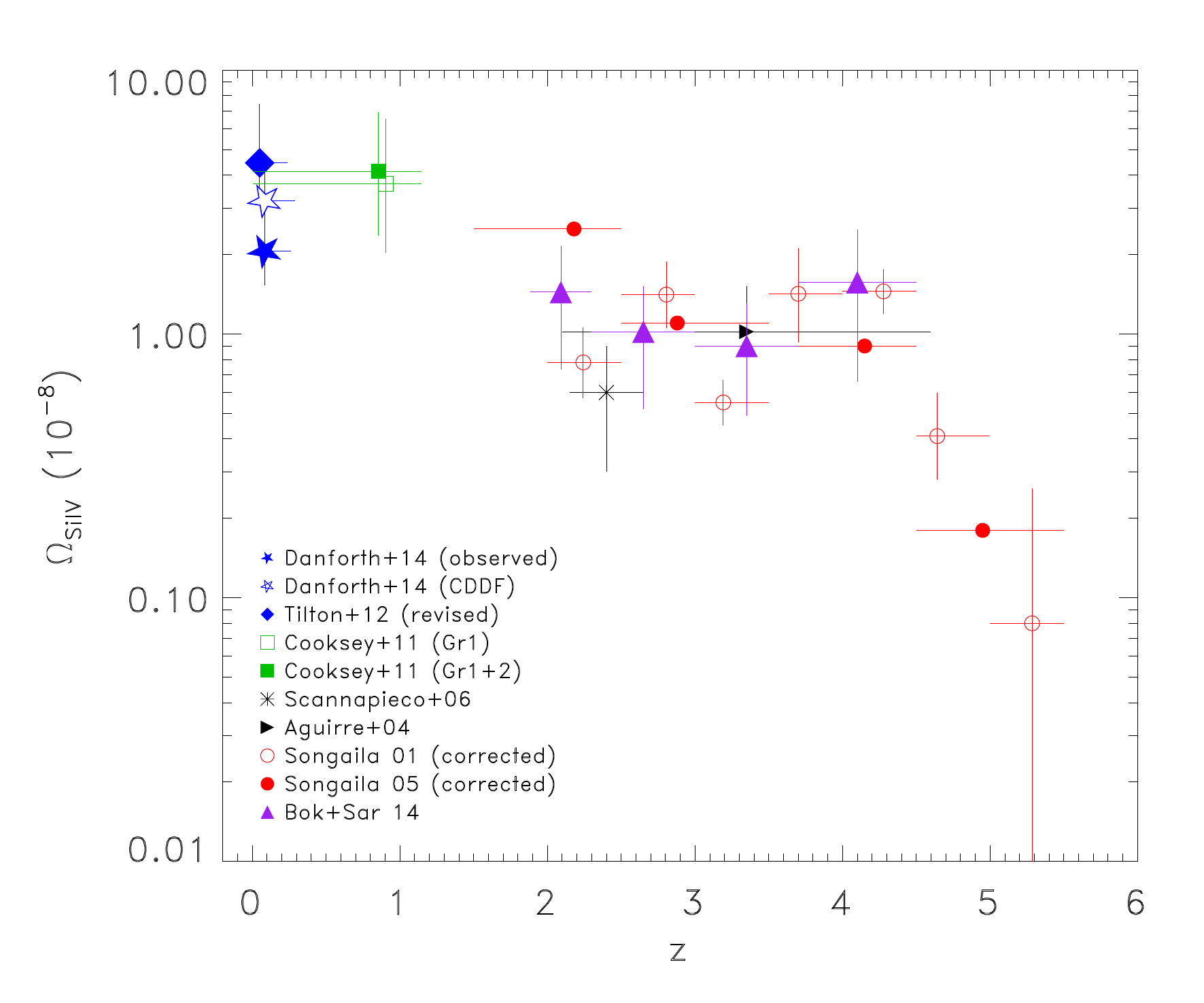}
\caption{\small Redshift evolution of \SiIV\ mass density, with $\Omega_{\rm SiIV}$ expressed relative to closure density
in units of $10^{-8} \, h_{70}^{-1}$.   Our HST/COS low-redshift measurement is shown together with values from 
past surveys with minimum column densities $\log N_{\rm SiIV}$ of 12.0  (Songaila 2001, 2005), 12.0 
(Scannapieco \etal\ 2006), 12.0  (Boksenberg \& Sargent 2014), and 12.5 (Danforth \etal\  2014). The \SiIV\  
abundance has increased by over a factor of 10 since $z \approx 5$.   }
\end{figure}



\begin{figure}
\epsscale{0.9}
 \plotone{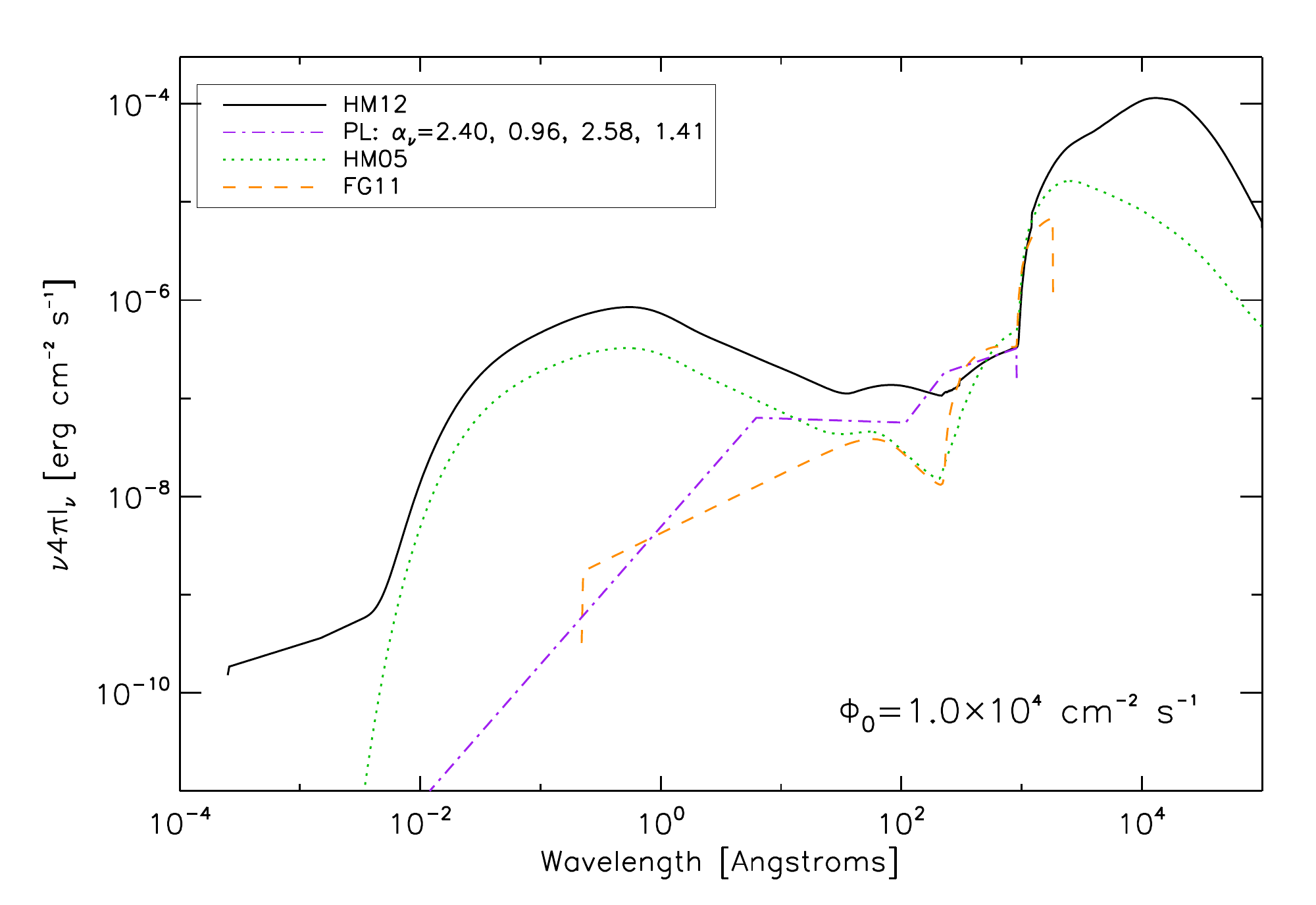} 
\caption{\small Several ionizing continua  at $z = 0$ are explored in our analysis of ionization corrections.  
These spectral energy distributions (SED) plot the monochromatic flux, $\nu F_{\nu} \equiv 4 \pi (\nu I_{\nu})$
from HM12 (Haardt \& Madau 2012), HM05 (unpublished 2005 tabulation from Haardt \& Madau 2001), and
FG11 (revised tabulation from Faucher-Gigu\`ere \etal\ 2009).  The PL is a broken power-law SED that
connects  the $F_{\nu} \propto \nu^{-1.41}$ AGN composite EUV spectrum (Shull \etal\ 2012) to the soft X-ray 
(1~keV) spectrum of AGN in the Lockman Hole (Hasinger 1994).  These SEDs are flux-normalized to common values
$\Phi_0 = 1.0 \times10^4$ phot~cm$^{-2}$~s$^{-1}$, the unidirectional, normally incident  photon flux (see 
Appendix B).  }
\end{figure}



\begin{figure}
 \epsscale{0.7} 
  \plotone{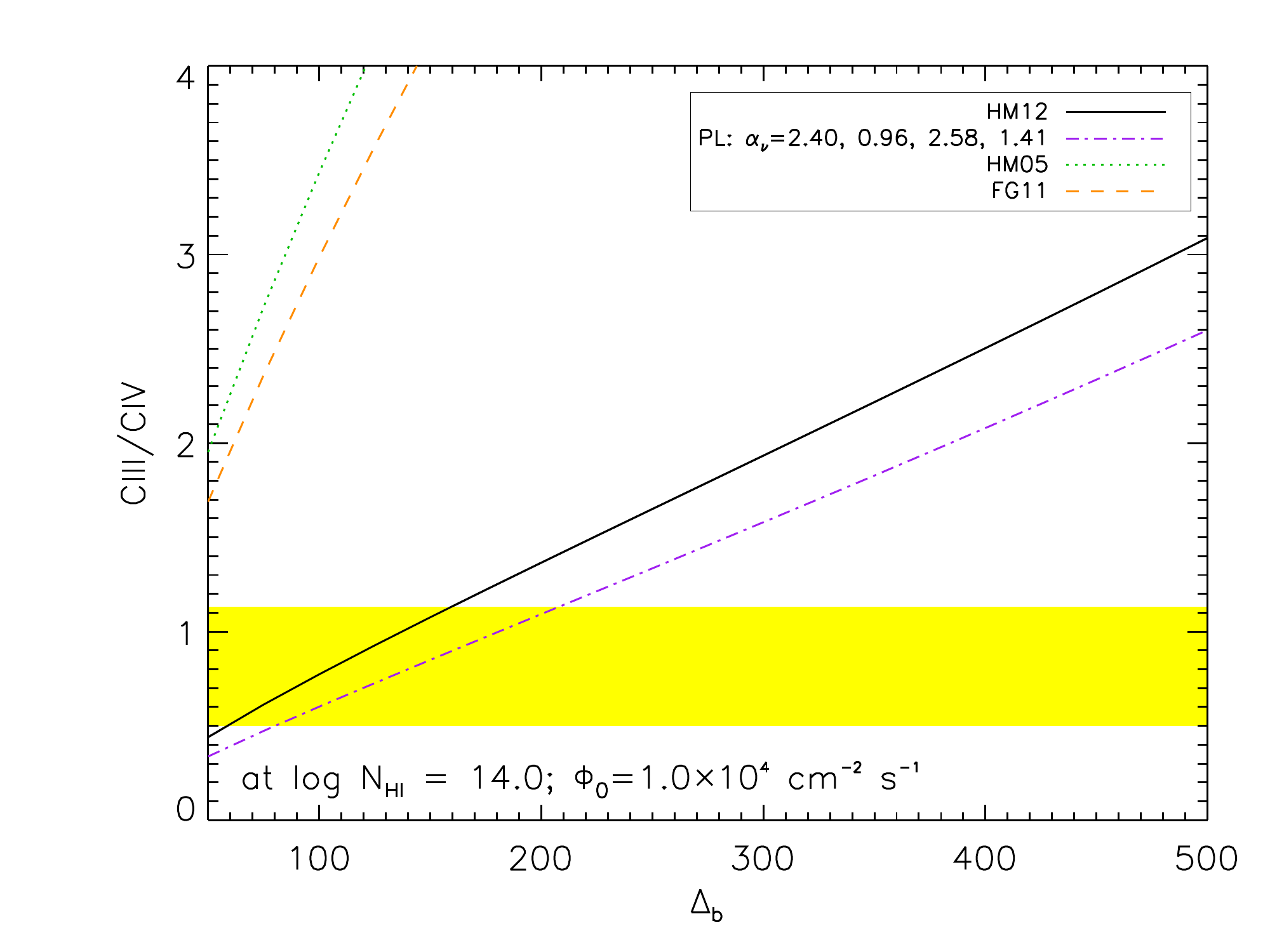} 
  \plotone{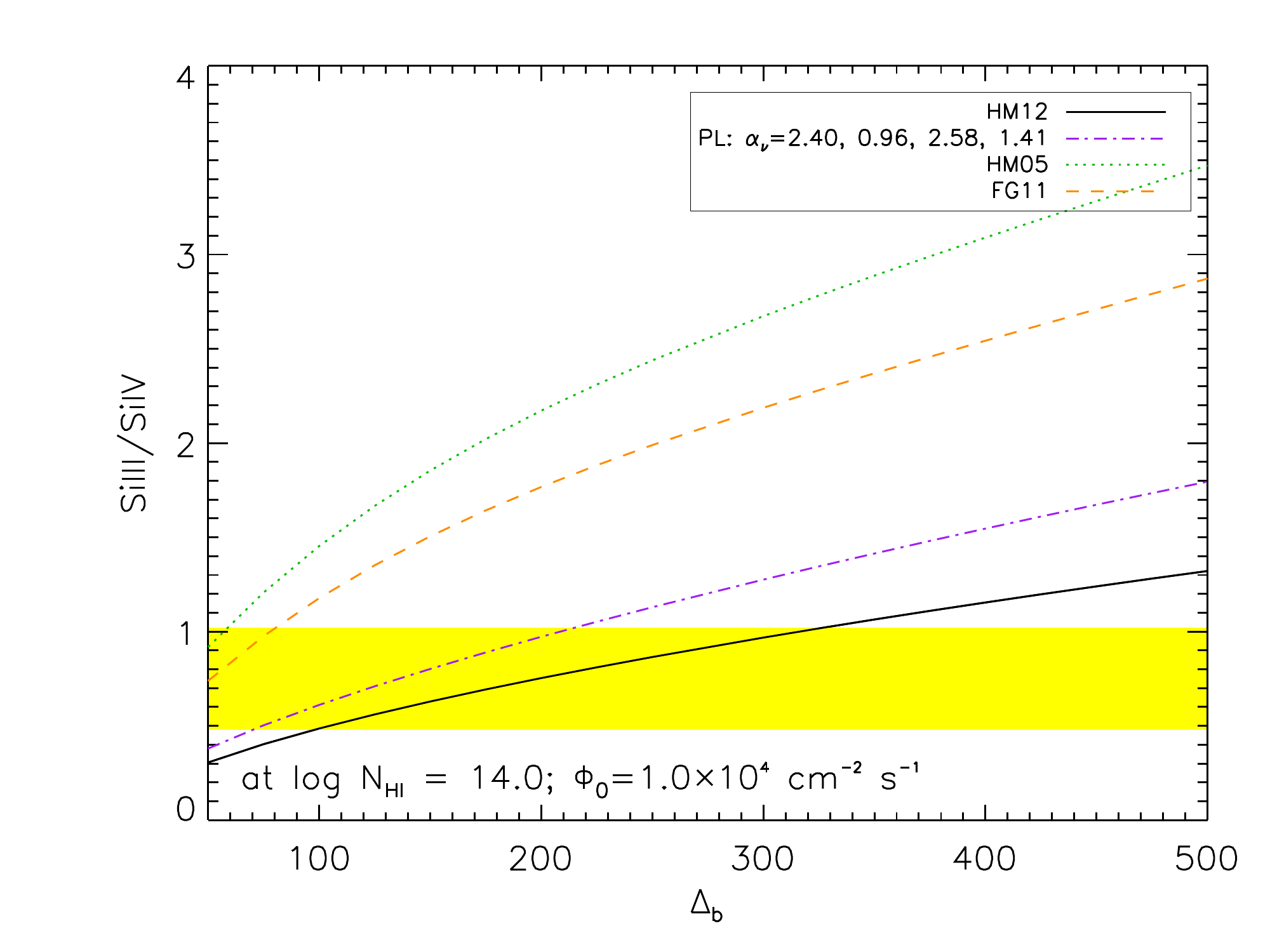} 
\caption{\small  Mean ion abundance ratios, \CIII/\CIV\ (top) and \SiIII/\SiIV\ (bottom) for five ionizing SEDs, 
labeled in box.  The top two curves (Haardt \& Madau 2005; Faucher-Gigu\`ere \etal\ 2009) yield somewhat 
higher ratios than observed (yellow bands), while the three lower curves (Haardt \& Madau \etal\ 2012,
and broken power law PL) give ratios consistent with our low-redshift COS observations, 
$\Omega_{\rm CIII} /  \Omega_{\rm CIV} = 0.70^{+0.43}_{-0.20}$ and  
$\Omega_{\rm SiIII} / \Omega_{\rm SiIV} = 0.67^{+0.35}_{-0.19}$.  
}
 \end{figure}



\begin{figure}
\epsscale{0.7}
\plotone{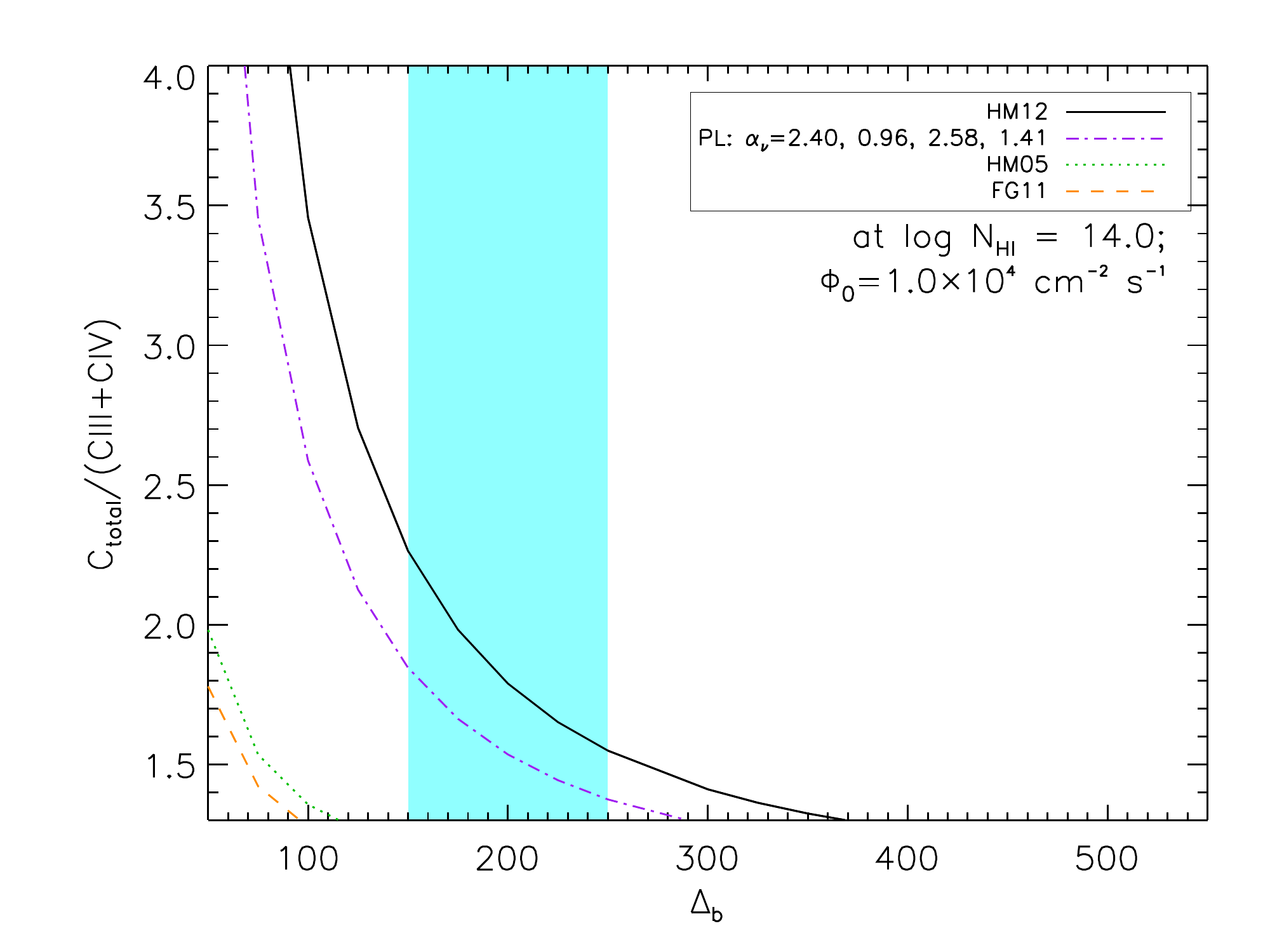} 
\plotone{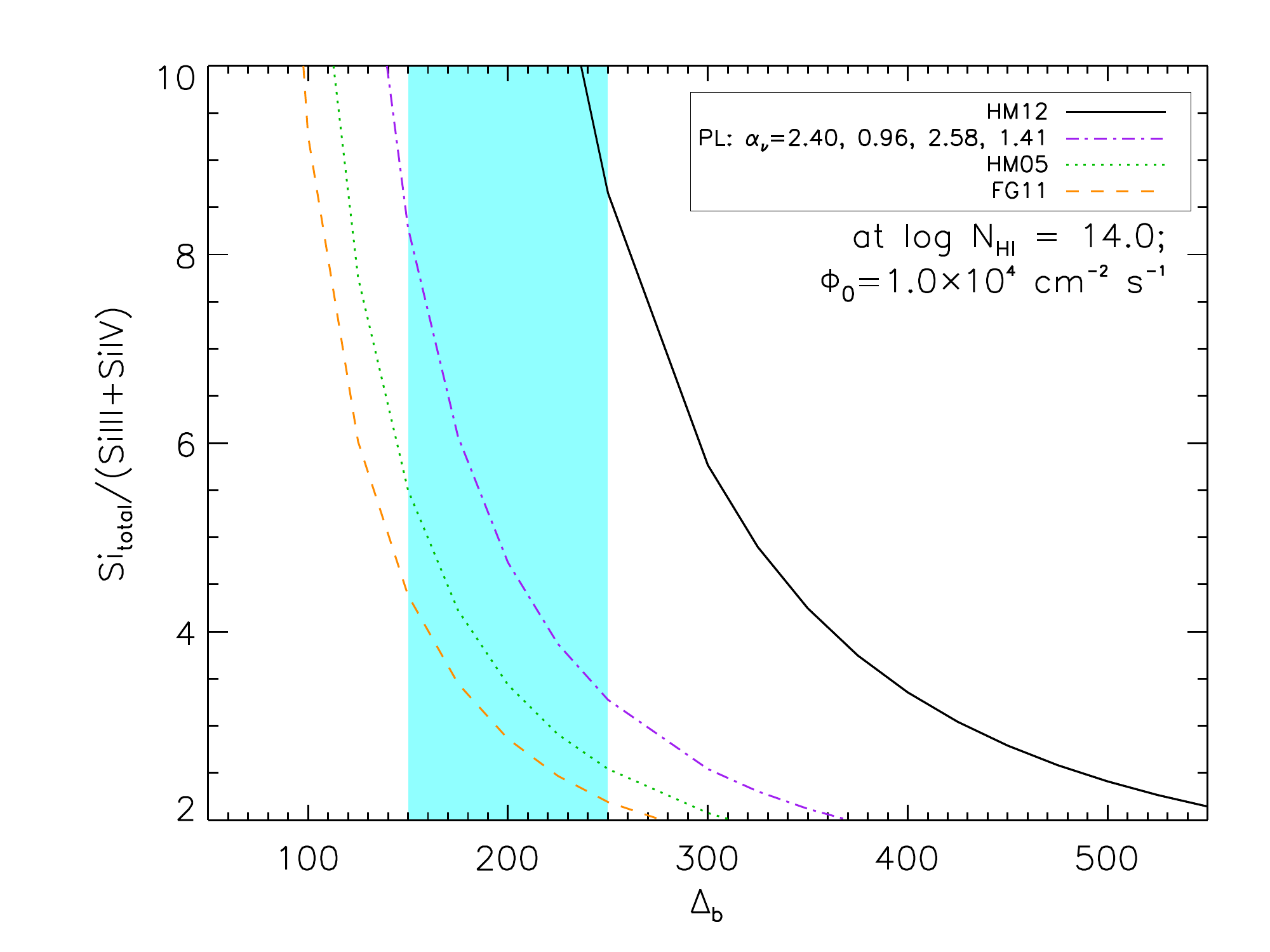} 
\caption{\small Ionization correction factors for total carbon and silicon relative to the two observed ion states,
(\CIII\ + \CIV) and (\SiIII\ + \SiIV), vs.\ baryon overdensity $\Delta_b$ . Curves correspond to ionizing continua 
labeled in boxes and plotted in Figure 3.   Vertical (blue) band shows the range of $\Delta_b = 200\pm50$
that provides consistent ionization corrections and metallicities for Si and C, with enhanced Si/C.  
See further discussion in Section 3.2.
 }
\end{figure}



\begin{figure}
 \epsscale{0.9} 
 \plotone{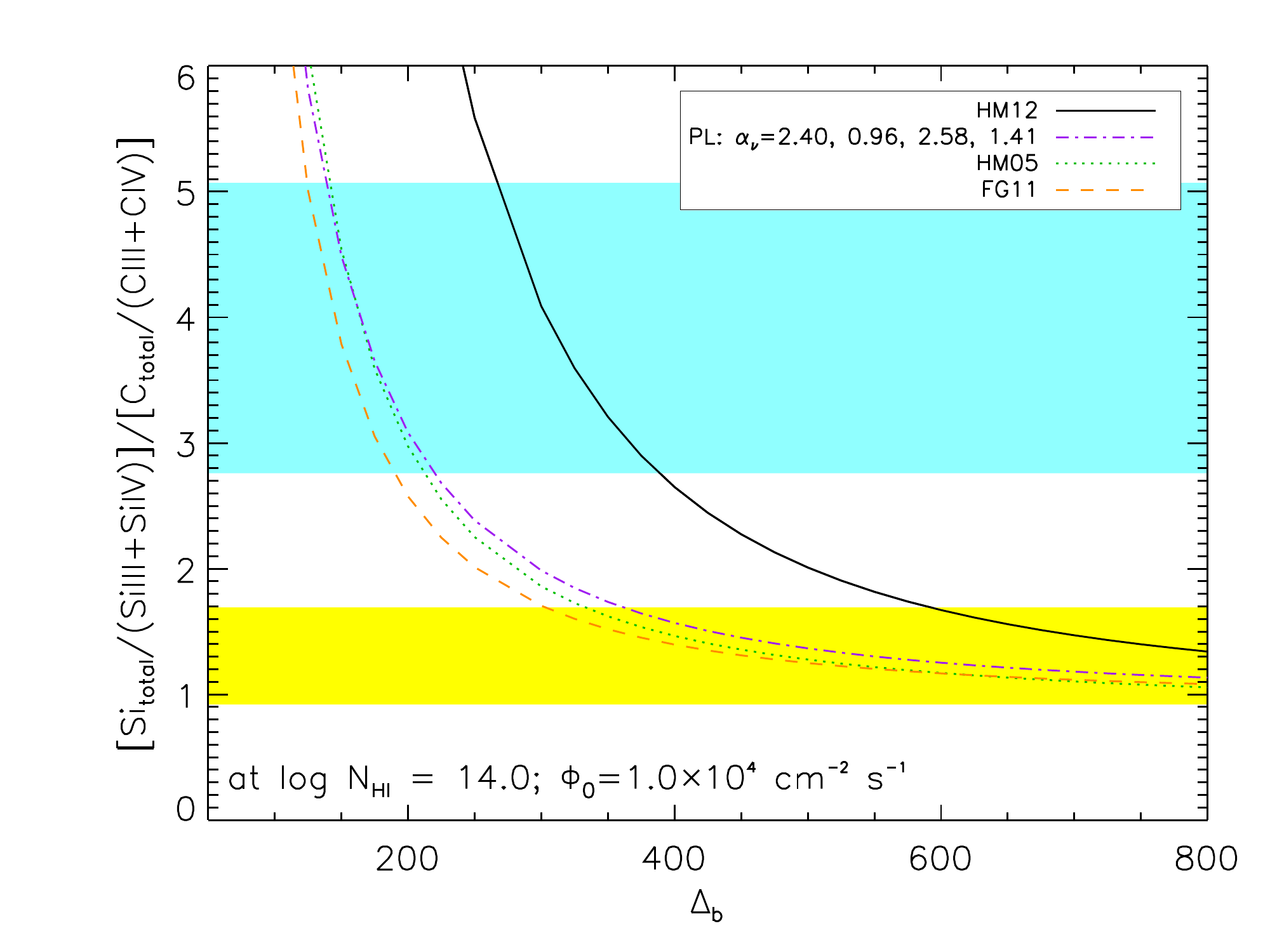} 
 \caption{\small  Ratio of the Si and C ionization correction factors,  ${\rm CF}_{\rm Si} / {\rm CF}_{\rm C}$,
 plotted vs.\ baryon overdensity $\Delta_b$ for several SEDs labeled in box.  Self-consistency requires that 
 the inferred metallicities agree with the Si/C mass-density ratio, which is 0.238 for solar abundances or 
 $3\times$ higher for alpha-enhanced Si/C.  This condition requires a ratio,
 ${\rm CF}_{\rm Si} / {\rm CF}_{\rm C}  = (1.17^{+0.52}_{-0.25}) \left[  (\rho_{\rm Si} / \rho_{\rm C}) / 0.238 \right]$,
 shown for Si/C in solar ratio (bottom band in yellow) and enhanced by factor of 3 (top band in blue). 
  }
\end{figure}



\begin{figure}
 \epsscale{0.85}
 \plotone{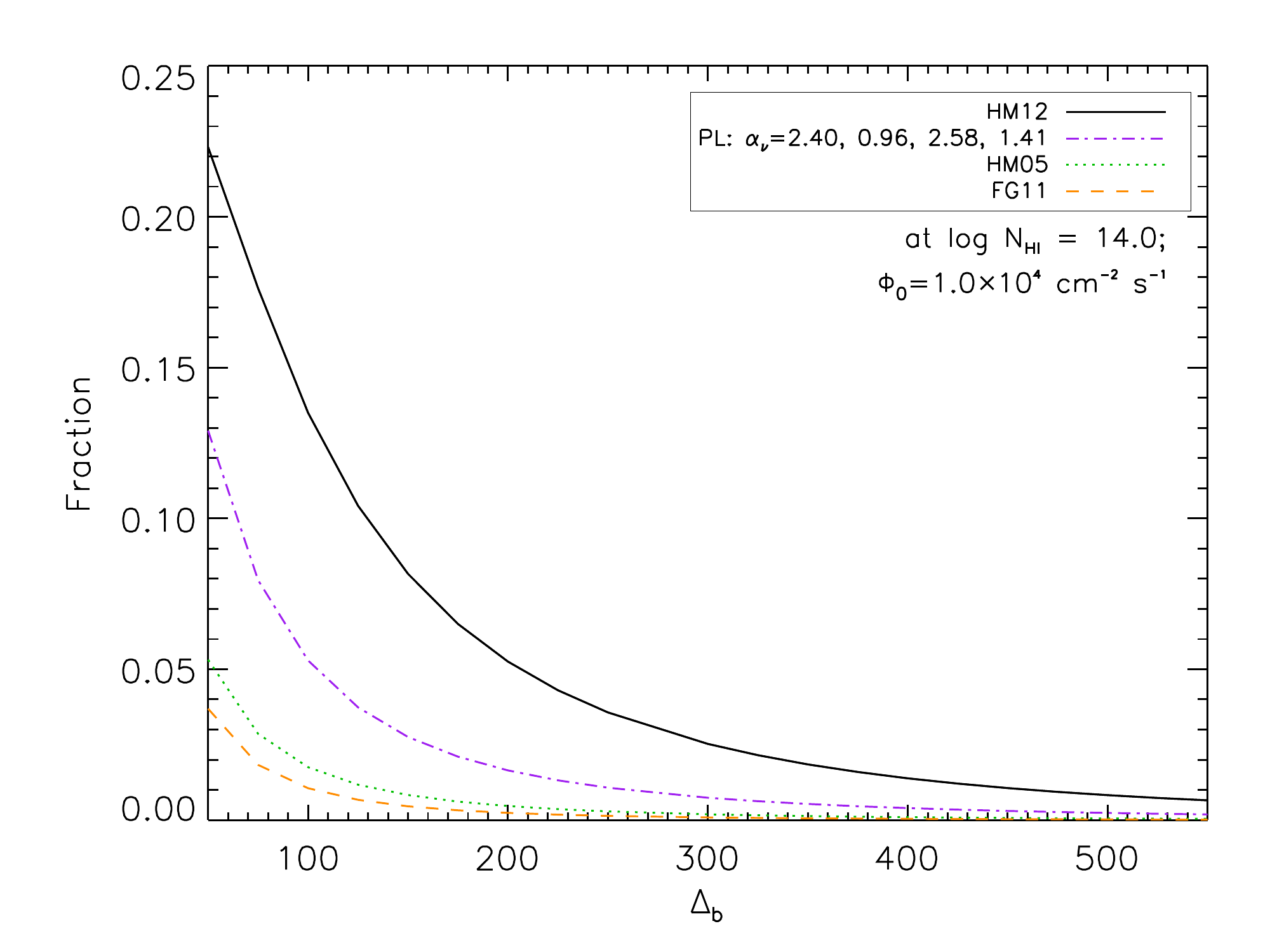} 
\caption{\small  Ionization fraction of \OVI\ vs.\ baryon overdensity $\Delta_b$, for several metagalactic radiation 
fields.  The top two curves show fraction for the full HM12 (Haardt \& Madau 2012) background, and removing 
photons above the carbon K-edge ($E \geq 290$~eV).   The three lower curves assume other backgrounds:
PL (broken power law); HM05 (Haardt \& Madau 2001); FG11 (Faucher-Gigu\`ere \etal\ 2009).   In the latter three
cases, the photoionized \OVI\ fraction is small at $\Delta_b \approx 200\pm50$.  }  
 \end{figure}


 \end{document}